\documentclass[fleqn,usenatbib]{mnras}

\usepackage{newtxtext,newtxmath}
\usepackage{booktabs}
\usepackage{orcidlink}
\usepackage{comment}
\usepackage[normalem]{ulem}
\usepackage[T1]{fontenc}

\newcommand{\EP}{\textit{Einstein Probe}}
\newcommand{\KW}{Konus-{\itshape Wind}}
\DeclareRobustCommand{\VAN}[3]{#2}
\let\VANthebibliography\thebibliography
\def\thebibliography{\DeclareRobustCommand{\VAN}[3]{##3}\VANthebibliography}

\usepackage{graphicx}	
\usepackage{amsmath}	
\usepackage[dvipsnames]{xcolor}
\usepackage[normalem]{ulem}

\definecolor{blazeorange}{rgb}{1.0, 0.4, 0.0}
\definecolor{seagreen}{rgb}{0.18, 0.55, 0.34}
\definecolor{rufous}{rgb}{0.66, 0.11, 0.03}
\definecolor{royalfuchsia}{rgb}{0.79, 0.17, 0.57}
\definecolor{scarlet}{rgb}{1.0, 0.13, 0.0}
\definecolor{royalpurple}{rgb}{0.47, 0.32, 0.66}
\definecolor{darkblue}{rgb}{0, 0, 0.66}
\definecolor{violet}{rgb}{0.5,0.,0.5}
\definecolor{nag}{RGB}{158, 56, 202}

\title[EP 260119a]{EP260119a: A High-Redshift Gamma-Ray Quiet Fast X-ray Transient Probing a Potentially Hidden Population of Relativistic Explosions}

\author[Angulo-Valdez et al.]{Camila Angulo-Valdez\,\orcidlink{0009-0002-6667-3294},$^{1}$\thanks{E-mail: camiangulo@astro.unam.mx (CAV)}, 
Asuka Kuwata\,\orcidlink{0000-0002-6169-2720},$^{2}$
Edilberto Aguilar-Ruiz\,\orcidlink{0000-0003-3502-4152},$^{2}$
Rosa~L.~Becerra\,\orcidlink{0000-0002-0216-3415},$^{1}$\thanks{E-mail: rbecerra@astro.unam.mx (RLB)},\newauthor
Ramandeep Gill\,\orcidlink{0000-0003-0516-2968},$^{2}$
Miguel~\'Angel Aloy\,\orcidlink{0000-0002-5552-7681},$^{3,4}$
Antonio de~Ugarte Postigo\,\orcidlink{0000-0001-7717-5085},$^{5}$
Alan~M.~Watson\,\orcidlink{0000-0002-2008-6927},$^{1}$,\newauthor
Chen-Wei Wang\,\orcidlink{0009-0008-8053-2985},$^{6}$
Diego~L\'opez-C\'amara\,\orcidlink{0000-0001-9512-4177},$^{7}$
Daniel A. Perley\,\orcidlink{0000-0001-8472-1996},$^{8}$
Aleksandra Bochenek\,\orcidlink{0009-0008-2714-2507},$^{8}$\newauthor
Simone~Dichiara\,\orcidlink{0000-0001-6849-1270},$^{9}$
Enrique~Moreno~M\'endez\,\orcidlink{0000-0002-5411-9352},$^{10}$
Nathaniel R. Butler\,\orcidlink{0000-0002-9110-6673},$^{11}$
Benjamin~Schneider\,\orcidlink{0000-0003-4876-7756},$^{5}$\newauthor
Jean-Luc~Atteia\,\orcidlink{0000-0001-7346-5114},$^{12}$
Stéphane~Basa\,\orcidlink{0000-0002-4291-333X},$^{5,13}$
William H.~Lee\,\orcidlink{0000-0002-2467-5673},$^{1}$
Dalya~Akl\,\orcidlink{0009-0006-4358-9929},$^{14,15}$
Sarah~Antier\,\orcidlink{0000-0002-7686-3334},$^{16}$\newauthor
Alexis Coleiro\,\orcidlink{0000-0003-0860-440X},$^{17}$
James~DeLaunay\,\orcidlink{0000-0001-5229-1995},$^{9}$
Damien~Dornic\,\orcidlink{0000-0001-5729-1468},$^{18}$
Jean-Grégoire~Ducoin\,\orcidlink{0009-0008-7341-4825},$^{18}$
Francis Fortin\,\orcidlink{0000-0003-3642-2267},$^{12}$\newauthor
Dmitry~Frederiks\,\orcidlink{0000-0002-1153-6340},$^{19}$
Leonardo~García-García\,\orcidlink{0000-0001-5125-1043},$^{20}$
Noémie~Globus\,\orcidlink{0000-0001-6148-6532},$^{20}$
Olivier Godet\,\orcidlink{0000-0001-7635-9544},$^{12}$\newauthor
Jamie Kennea\,\orcidlink{0000-0002-6745-4790},$^{9}$
Gianluca~Lombardi\,\orcidlink{0000-0003-3412-0556},$^{21,22}$
Alexandra L.~Lysenko\,\orcidlink{0000-0002-3942-8341},$^{19}$
Francesco~Magnani\,\orcidlink{0009-0000-6101-7373},$^{18}$\newauthor
Tyler Parsotan\,\orcidlink{0000-0002-4299-2517},$^{23}$
Margarita~Pereyra\,\orcidlink{0000-0001-6148-6532},$^{20,24}$
Anna~Ridnaia\,\orcidlink{0000-0001-9477-5437},$^{19}$
Samuele Ronchini\,\orcidlink{0000-0003-0020-687X},$^{25,26}$\newauthor
Fredd~Sánchez-Álvarez\,\orcidlink{0009-0009-5612-3759},$^{1}$
Dmitry~Svinkin\,\orcidlink{0000-0002-2208-2196},$^{19}$
Anastasia~Tsvetkova\,\orcidlink{0000-0003-0292-6221},$^{19}$\newauthor
Mikhail~Ulanov\,\orcidlink{0000-0002-0076-5228},$^{19}$
Peter~Veres\,\orcidlink{0000-0002-2149-9846},$^{27}$
Hui Yang\,\orcidlink{0000-0002-8832-6077},$^{12}$
\\
$^{1}$ Universidad Nacional Aut\'onoma de M\'exico. Instituto de Astronom\'ia. A.P. 70-264, 04510. Ciudad de M\'exico, M\'exico.\\
$^{2}$ Instituto de Radioastronomía y Astrof\'isica, Universidad Nacional Aut\'onoma de M\'exico, Antigua Carretera a P\'atzcuaro \# 8701,\\
Ex-Hda. San Jos\'e de la Huerta, Morelia, Michoac\'an, M\'exico C.P. 58089, M\'exico\\
$^{3}$ Departament d'Astronom\'{\i}a i Astrof\'{\i}sica, Universitat de València, 46100 Burjassot, Spain\\
$^{4}$ Observatori Astronòmic, Universitat de València, 46980 Paterna, Spain\\
$^{5}$ Aix Marseille Univ., CNRS, CNES, LAM, Marseille, France\\
$^{6}$ State Key Laboratory of Particle Astrophysics, Institute of High Energy Physics, Chinese Academy of Sciences, Beijing 100049, People’s Republic of China\\
$^{7}$ Instituto de Ciencias Nucleares, Universidad Nacional Aut\'onoma de M\'exico, Apartado Postal 70-264, 04510 M\'exico, CDMX, M\'exico\\
$^{8}$ Astrophysics Research Institute, Liverpool John Moores University, 146 Brownlow Hill, Liverpool L3 5RF, UK\\
$^{9}$ Department of Astronomy and Astrophysics, The Pennsylvania State University, 525 Davey Lab, University Park, PA 16802, USA\\
$^{10}$ Facultad de Ciencias, Universidad Nacional Aut\'onoma de M\'exico, Apartado Postal 70-542, 04510 M\'exico, CDMX, M\'exico\\
$^{11}$ School of Earth and Space Exploration, Arizona State University, Tempe, AZ 85287, USA\\
$^{12}$ IRAP, Université de Toulouse/CNRS/CNES, 9 avenue du colonel Roche, 31028 Toulouse, France\\
$^{13}$ Aix-Marseille University, CNRS, OSU-Pytheas, France\\
$^{14}$ New York University Abu Dhabi, PO Box 129188, Saadiyat Island, Abu Dhabi, UAE\\
$^{15}$ Center for Astrophysics and Space Science (CASS), New York University Abu Dhabi, Saadiyat Island, PO Box 129188, Abu Dhabi, UAE\\
$^{16}$ IJCLAB, Université Paris Saclay, Orsay, France\\
$^{17}$ Université Paris Cité, CNRS, CEA, Astroparticule et Cosmologie, F-75013 Paris, France\\
$^{18}$ Aix Marseille University, CNRS, CPPM, Marseille, France\\
$^{19}$ Ioffe Institute, 26 Politekhnicheskaya, St. Petersburg, 194021, Russia\\
$^{20}$ Instituto de Astronom{\'\i}a, Universidad Nacional Aut\'onoma de M\'exico, km 107 Carretera Tijuana-Ensenada, 22860 Ensenada, Baja California, México\\
$^{21}$ GRANTECAN S.A., Cuesta de San Jos\'e s/n, E-38712 Bre\~na Baja, La Palma, Spain\\
$^{22}$ Instituto de Astrofísica de Canarias, E-38205 La Laguna, Tenerife, Spain\\
$^{23}$ Astrophysics Science Division, NASA Goddard Space Flight Center, Greenbelt, MD 20771, USA\\
$^{24}$ Secretar\'ia de Ciencia, Humanidades, Tecnolog\'ia, e Innovaci\'on\\
$^{25}$ Gran Sasso Science Institute (GSSI), I-67100 L’Aquila, Italy\\
$^{26}$ INFN, Laboratori Nazionali del Gran Sasso, I-67100 Assergi, Italy\\
$^{27}$ Department of Space Science, University of Alabama in Huntsville, Huntsville, AL 35899, USA\\
}

\date{Accepted XXX. Received YYY; in original form ZZZ}

\pubyear{\the\year{}}

\begin{document}
\label{firstpage}
\pagerange{\pageref{firstpage}-\pageref{lastpage}}
\maketitle
\clearpage

\begin{abstract}
Gamma-ray quiet fast X-ray transients (FXTs) provide a new approach for studying relativistic explosions that evade traditional gamma-ray triggers. In this work, we present multi-wavelength observations and analysis of EP260119a, a high-$z$ FXT detected by \textit{Einstein Probe}/WXT and followed up with COLIBRÍ, the Liverpool Telescope, and other facilities. Spectroscopy yields a redshift of $z = 5.47$, making EP260119a the most distant FXT detected by \textit{Einstein Probe} to date.
Despite its luminous X-ray and optical emission, no prompt gamma-ray counterpart was detected by \textit{SVOM}/ECLAIRs, \textit{SVOM}/GRM, \textit{Swift}/BAT, or \textit{Konus}/Wind, despite contemporaneous coverage. The broadband afterglow is well described by synchrotron emission from a uniform relativistic jet propagating into a shallowly stratified external medium close to the constant-density limit, indicating a standard relativistic explosion despite the absence of detectable gamma rays. 
Occupying the extreme high-$z$ end of the growing \textit{Einstein Probe} FXT sample, EP260119a supports the possibility that soft X-ray surveys are uncovering relativistic transients that remain undetected by current gamma-ray instruments.
Our results indicate that some explosions in the early Universe escape conventional gamma-ray surveys and demonstrate the value of combining sensitive soft X-ray discovery with rapid optical follow-up to obtain a more complete view of the high-$z$ transient population.
\end{abstract}

\begin{keywords}
X-rays: bursts - (transients:) gamma-ray bursts - (stars:) gamma-ray burst: general
\end{keywords}

\section{Introduction}

Fast X-ray transients (FXTs) are extragalactic high-energy transients characterised by bright X-ray emission in the 0.5–10 keV band, with durations ranging from tens to thousands of seconds, and little or no detected gamma-ray emission \citep{Yuan2025,Quirola2024}. They occupy a specific region in terms of spectral softness, X-ray-to-gamma-ray fluence ratio, and prompt duration, different from classical gamma-ray bursts (GRBs) and the softest X-ray flashes (XRFs) \citep{Heise2003,Sakamoto2005}, but their physical nature remains under debate.

For decades, FXTs were detected only sporadically. Missions such as \textit{BATSE} \citep{Fishman1992}, BeppoSAX \citep{Boella1997}, \textit{HETE-2} \citep{Ricker2003}, \textit{Swift} \citep{Gehrels2004}, \textit{Chandra} \citep{Weisskopf2000}, \textit{NuSTAR} \citep{Harrison2013}, and \textit{MAXI} \citep{Matsuoka2009} reported isolated soft, short-lived X-ray outbursts, frequently classified as X-ray flashes (XRFs) or as sub-energetic, GRB-like events with faint or absent $\gamma$-ray emission \citep[see e.g.][]{Kippen2001,Heise2003,Stratta2007,Sakamoto2005,Liang2007}.

This discovery history left FXTs in an ambiguous relationship with the broader family of high-energy transients. In the standard picture, XRFs and X-ray rich GRBs form a spectral continuum with classical GRBs, differing primarily in their spectral peak energy, $E_\mathrm{peak}$, which decreases from the MeV regime in hard GRBs to a few keV in the softest XRFs \citep{Kippen2001,Sakamoto2005}. 
FXTs appear to occupy an intermediate place of this distribution, and in some cases extend beyond it. While several FXTs exhibit afterglow properties similar to those of GRBs, others are remarkably soft and entirely gamma-ray quiet. Consequently, it remains unclear whether FXTs represent the low-$E_\mathrm{peak}$ tail of the GRB population or a physically distinct class of relativistic transients \citep{Yang2019, Quirola2022, Quirola2023, Quirola2024, Lansbury2017, Levan2014}.

The \textit{Einstein Probe} (\textit{EP}) mission \citep{Yuan2025,Yuan2022} has transformed the study of FXTs. Equipped with the Wide-field X-ray Telescope (WXT), which combines an instantaneous field of view of 3600 deg$^2$ with a sensitivity of approximately 1 mCrab in 1000~s in the 0.5–4~keV band \citep{Cheng2025}, \textit{EP} is discovering FXTs at an unprecedented rate. Prior to the launch of \textit{EP} in 2024, only $\sim40$ FXTs had been identified through two decades of archival searches with previous X-ray missions \citep{Quirola2022,Quirola2023,Quirola2024}, with a detection rate of $\approx 2$~events~yr$^{-1}$. In its first two years of observations (to the end of June 2026), however, \textit{EP} had reported through GCNs, around 200 extragalactic events, or $\approx 80$~events~yr$^{-1}$, increasing the discovery rate by more than an order of magnitude and, for the first time, enabling statistically meaningful population studies of FXTs. Moreover, its sensitivity in the soft X-ray band makes \textit{EP} particularly well suited to detecting high-redshift transients, opening a new window onto the distant transient Universe.

The growing sample of FXTs discovered by {\EP} has begun to clarify the nature of these events and demonstrates why they have become central to transient astrophysics. A fraction are unambiguously GRB-related, such as EP240315a, associated with a long GRB at $z = 4.859$ 
\citep{Liu2025, Gillanders2024,Levan2025}, and the redshift distribution of spectroscopically confirmed EP transients is statistically consistent with that of long GRBs \citep{OConnor2025}. Others, such as EP240408a, EP240414a, and EP241021a, lack detectable gamma-ray emission and span a variety of timescales and luminosities that are difficult to associate with a single progenitor type \citep{Zhang2025,Shu2025}. Establishing where each FXT falls between these extremes, and understanding the progenitors, emission mechanisms, and cosmic distribution of these soft X-ray explosions is now an important goal of the field.

Among the emerging FXT population, those lacking detected gamma-ray emission, hereafter called gamma-ray quiet FXTs (GQFXTs), and residing at high-$z$ are particularly revealing. GQFXTs may arise from intrinsically soft prompt spectra or suppressed gamma-ray production \citep[e.g.,][]{Becerra2026, Sun2025, Yadav2025, Gianfagna2025, Jonker2026, Quirola2026}.

Throughout this work, we use the term gamma-ray quiet in an observational sense, referring to FXTs for which no contemporaneous gamma-ray counterpart was detected despite suitable instrumental coverage. This classification does not by itself imply that the intrinsic gamma-ray emission was absent or substantially weaker than that of classical GRBs, since such an inference depends on the unknown prompt spectral shape and on the sensitivity of the available instruments.

Their discovery at $z\gtrsim3$ demonstrates that this population extends to the same cosmological distances as classical long GRBs \citep[e.g.,][]{Liu2025}, providing an opportunity to investigate relativistic explosions during the first billion years of cosmic history and to assess whether these events share a similar physical origin as the broader GRB population.

Against this observational and theoretical background, EP260119a, detected on the 19th of January 2026 \citep{43447}, represents an important example within the emerging population of high-$z$ GQFXTs. The event was identified by EP/WXT as a soft X-ray transient with X-ray emission detected over a time interval of approximately 300~s \citep{43447}, and subsequently associated with an extremely red optical afterglow. Spectroscopic observations confirmed a redshift of $z=5.47$ \citep{43469}, placing it at $\sim1$~Gyr after the Big Bang and making it the highest-$z$ FXT detected by {\EP} to date. Less than 20 relativistic transients of any kind have been found at $z \geq 5$, most of them long GRBs \citep{Dainotti2024}; EP260119a extends this sample of distant relativistic explosions where before only prompt gamma-ray triggers had been detected.

In this work, we exploit the combination of the extreme distance of EP260119a, the stringent gamma-ray upper limits, and the multi-wavelength follow-up campaign to investigate whether this high-$z$ GQFXT is consistent with a standard relativistic explosion. We further place EP260119a in the context of the emerging GQFXT population and discuss its implications for the origin of gamma-ray quiet transients in the early Universe.

The manuscript is organised as follows. In Section~\ref{sec:observations} we describe the X-ray and optical observations and in Section~\ref{sec:gammaray}, we summarise the upper limits provided by {\KW} \citep{Aptekar1995}, \textit{SVOM} (through the ECLAIRs and GRM instruments) \citep{Wei2016}, and \textit{Swift}/BAT, as well as the physical implications of the non gamma-ray detection. We discuss the possible host galaxy in Section~\ref{sec:hg}. In Section~\ref{sec:evolution} we present and discuss the modelling used to describe the dataset. We further compare the properties of EP260119a with samples of GRBs and FXTs and the implications of our results in Section~\ref{sec:discussion}. Finally, we summarise our conclusions in Section~\ref{sec:summary}.

Throughout this work, we adopt a flat $\Lambda$CDM cosmology with $H_0 = 67.7$~$\mathrm{km\ s^{-1}\ Mpc^{-1}}$ and $\Omega_m = 0.31$ \citep{Planck2020}. 

\section{Observations}
\label{sec:observations}
\subsection{Detection in X-rays: \EP}
\label{sec:xray-obs}

On January 19, 2026, at 00:13:48 UTC ($T_0$), the EP/WXT triggered on EP260119a \citep{43447}. The automated localization yielded a position of 10:32:19.7 +65:30:18 (J2000)
with a 90\% confidence error radius of 2.3\arcmin, including systematic uncertainties. A follow-up pointing by the EP/FXT, commencing about 7.5~hours after trigger, revealed an uncatalogued X-ray source at 10:32:25.13 +65:29:51.0 (J2000) with an uncertainty of 20\arcsec \citep{43449}. This position was later refined to 10:32:25.13 +65:29:51.00 (J2000) with an uncertainty of 10\arcsec \citep{43465}.

\begin{figure}
	\includegraphics[clip, width=0.9\linewidth]{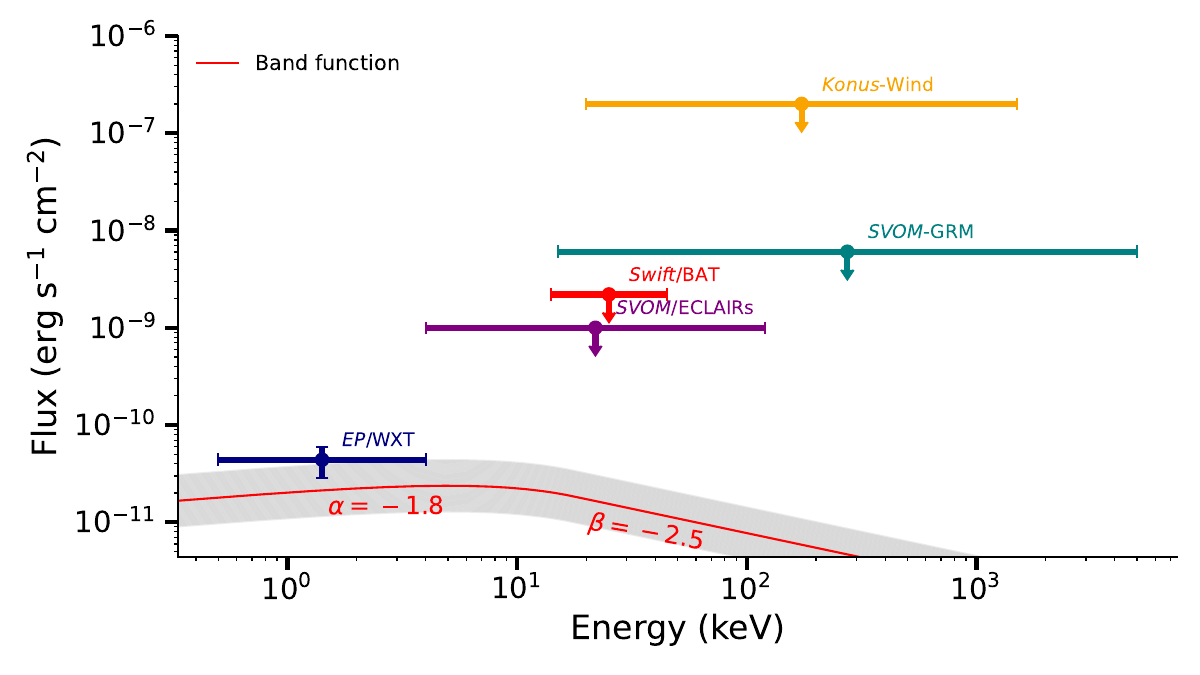}
    \caption{Spectral energy distribution of EP260119a at high energies. The EP/WXT flux and the $3\sigma$ upper limits derived from \emph{SVOM}/ECLAIRs, \emph{SVOM}/GRM, \textit{Swift}/BAT are shown together with the \textit{Konus-Wind} 90\% confidence peak-flux upper limit in the 20--1500~keV band, evaluated on a 2.944~s timescale assuming a typical long-GRB spectrum. A representative Band-function spectrum consistent with the EP/WXT measurement and the current gamma-ray upper limits is overplotted. The corresponding Band-function spectrum has a peak energy of $E_{\rm peak}=33.1^{+8.34}_{-6.76}$\,keV in the GRB rest frame. 
    }
    \label{fig:sedhe}
\end{figure}

\cite{43447} report that the WXT light curve shows a transient lasting $\sim$300~s and that the time-averaged spectrum in the 0.5-4~keV band can be described by an absorbed power law with photon index $\Gamma =1.8 \pm 0.6$ (with Galactic absorption fixed at $N_{\rm H} = 1.2\times10^{20}$~cm$^{-2}$), corresponding to an average unabsorbed flux of $(4.4^{+1.8}_{-1.3}) \times 10^{-11}$~erg~s$^{-1}$~cm$^{-2}$. The corresponding flux is shown in Figure~\ref{fig:sedhe}.

In its analysis of the afterglow, the EP/FXT team reported in \cite{43465} that the X-ray spectrum can be fitted with an absorbed power law ($N(E) \propto E^{-\Gamma}$) with photon index $\Gamma = 1.99 \pm 0.13$ (with a Galactic absorption value fixed at $N_{\rm H} = 1.20 \times 10^{20}$~cm$^{-2}$), and a derived unabsorbed flux in the 0.5-10 keV range of $\sim (5.69^{+0.60}_{-0.55}) \times 10^{-13}$~erg~s$^{-1}$~cm$^{-2}$.

\subsection{Discovery and follow-up of the optical counterpart}

\subsubsection{COLIBRÍ}

\begin{figure}
\centering
	\includegraphics[clip,width=0.8\linewidth]{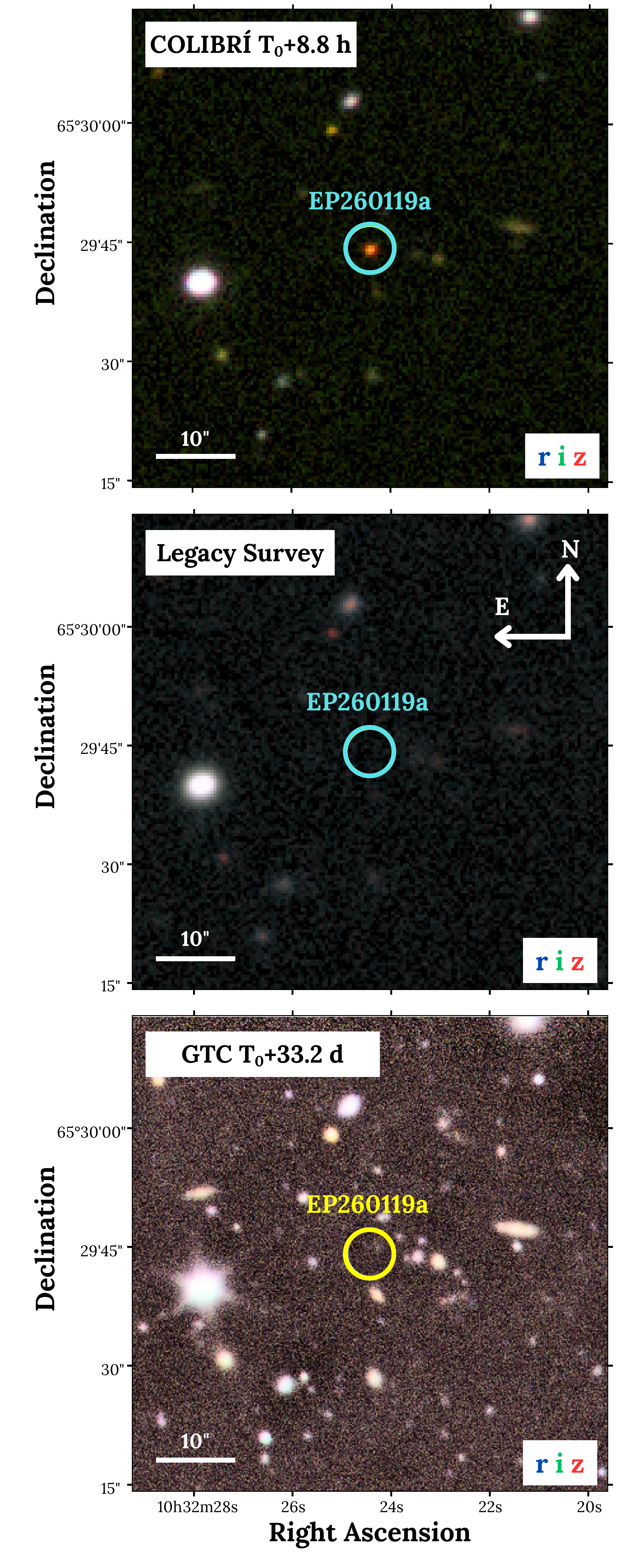}
    \caption{We show the false colour image obtained with COLIBRÍ in $riz$ in our first night. The images started at $T_0+8.8$~hours with a total exposure of 53 minutes (top panel) in comparison with the same field in the DESI Legacy Survey DR10 (middle panel) \citep{Dey2019} and the late image acquired with the GTC/HIPERCAM instrument. \label{fig:field}}
\end{figure} 

We observed the field of EP260119a with the DDRAGO wide-field imager on the COLIBRÍ telescope. COLIBRÍ\footnote{\url{https://www.colibri-obs.org/}} is a Franco-Mexican, fast, robotic 1.3~m telescope operated by the Observatorio Astronómico Nacional in the Sierra de San Pedro Mártir, Baja California \citep{Basa2022,Basa2026}. DDRAGO is a two-channel imager, with the blue channel working in $gri$ and the red channel in $zy$ \citep{Langarica2024}. Our observations used the $r$, $i$, $z$, and $y$ filters, which closely match the SDSS/Pan-STARRS filter system \citep[see more details in][]{Angulo2026}. 

The COLIBRÍ control system responds to EP automated GCN notices and automatically schedules them for observations, but for this event, no automated notice was issued by EP due to the low SNR \citep{43447}. However, as part of our routine science monitoring, our team read this GCN Circular and manually programmed the event. Our observations began at 09:00 UTC ($T_0+8.78$ hours).

We observed the event with COLIBRÍ over 3 nights. On the night of January 19, 2026, we identified an uncatalogued optical counterpart consistent with the EP/FXT localisation with RA, Dec = 10:32:24.4, +65:29:44.2 with an uncertainty of $0.5$~arcsec and preliminary magnitudes of $r = 23.02 \pm 0.14$ and $z = 20.08 \pm 0.03$. We reported these values in \cite{43450}. Using our images of this night, we illustrate the field of EP260119a in Figure~\ref{fig:field}, comparing with the same field in the DESI Legacy Survey catalog \citep{Dey2019}. Noting the brightness in $z$ and the very red colour of $r-z \approx 3$, we then performed photometry in the \emph{rizy} bands to obtain a photometric redshift for the transient. On the nights of January 20 and 21 (UTC), 2026, we observed in the \emph{i} and \emph{z} bands only, due to the faintness of the source in $r$. 

In our refined analysis, we adjusted the position to RA, Dec = 10:32:24.416, +65:29:44.10

We reduced, coadded, calibrated, and analysed our observations with the COLIBRÍ ASU pipeline (see Appendix~\ref{app:colibriasu}). The photometry was calibrated using nearby stars from the PanSTARRS DR1 catalog \citep{Flewelling2020,Magnier2020}. Photometry for the first night of observations is binned in groups of 10, 11, and 20 single exposures for the $z$, $i$, and $r$ filters, respectively. Our photometry (not corrected for Galactic extinction) is listed in Table~\ref{tab:photometry} and is shown in Figure~\ref{fig:LC}. 

All the magnitudes presented in this work are in the AB system without the application of any colour terms and are not corrected for the Galactic extinction in the direction of the burst $E_{(B-V)}=0.01$ \citep{Schlafly2011}, which implies extinctions of $A_r=0.02$, $A_i=0.02$, $A_z=0.01$, and $A_y=0.01$. In our analysis, we apply this correction. 

\subsubsection{Liverpool Telescope}

We observed the field of EP260119a with the IO:O telescope on the Liverpool Telescope \citep{Steele2004} starting at 23:58 UTC on the 19th of January 2026 ($T_0 + 23.75$ hours). We obtained $12 \times 100$ seconds in the SDSS $r$ and $i$ filters and $11 \times 100$ seconds in SDSS $z$ filter. We detected the optical transient and reported preliminary magnitudes in \cite{43472}.

\begin{figure}
	\includegraphics[clip,width=\linewidth]{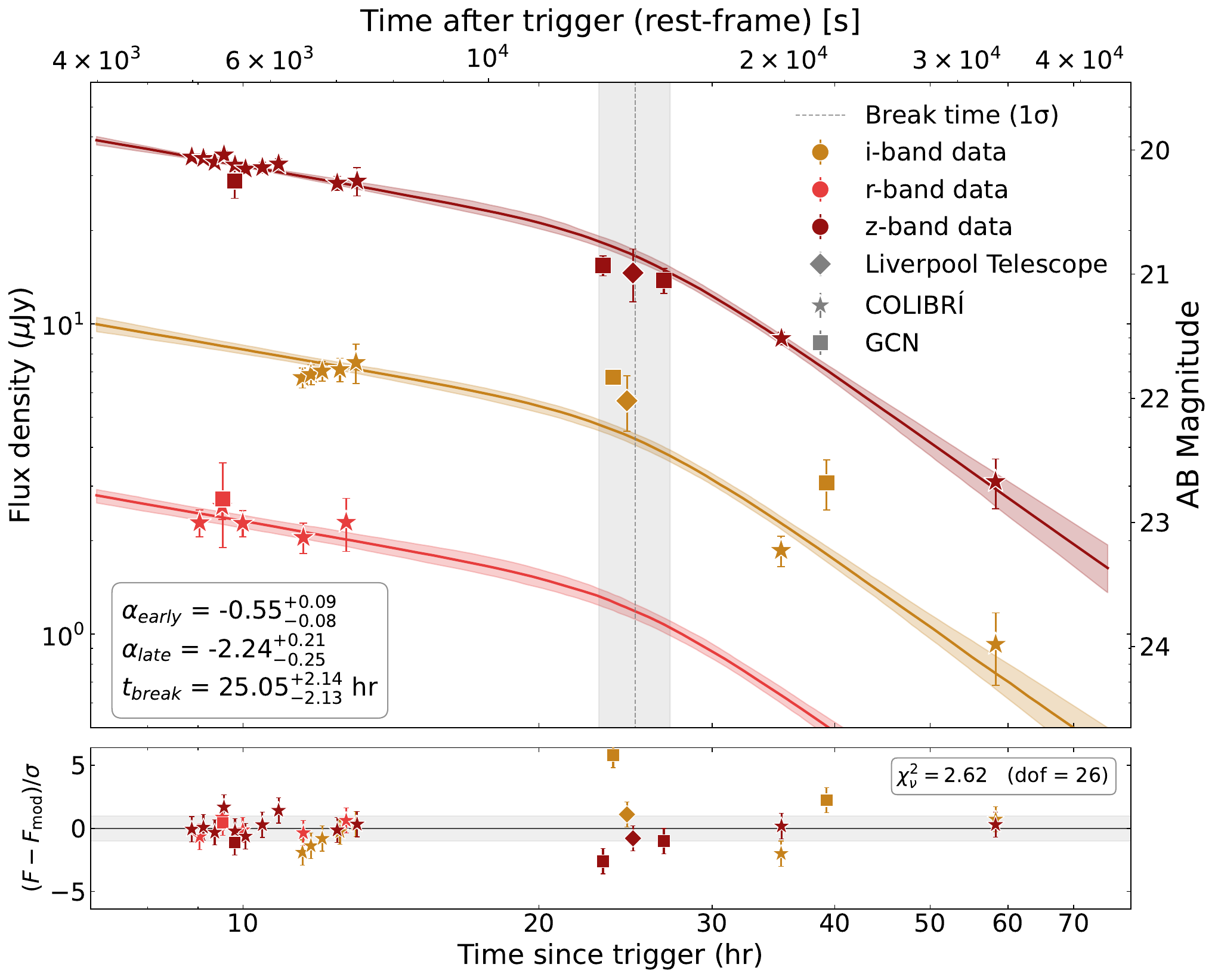}
    \caption{\textbf{\emph{Top:}} We show the optical light curve obtained with COLIBRÍ (stars) in $riz$, Liverpool Telescope (diamonds) in $riz$, and the ones taken from the General Circular Network (squares), corrected for Galactic extinction.  We also show the temporal decays for the plateau and the steep decay seen before and after $t_{\rm{break}} = T_0 + 25.05^{+2.14}_{-2.13}$~h. \textbf{\emph{Bottom:}} Residuals $(F - F_{\rm{model}})/\sigma$ evaluated at the posterior-median model for each data point, with the shaded band marking the $\pm 1\sigma$ region. The reduced $\chi^2 = 2.62$ of the fit and its degrees of freedom (dof = 26) are shown in the figure.
    \label{fig:LC}}
\end{figure}

The images were stacked onto a single image of 1200~s of exposure for the $r$ and $i$ filters, and of 1100~s of exposure for the $z$ filter. The observations were analysed with the COLIBRÍ ASU pipeline consistently to the COLIBRÍ data, using nearby stars from PanSTARRS DR1 catalog \citep{Flewelling2020,Magnier2020} for calibration. Our photometry from the Liverpool Telescope is listed in Table~\ref{tab:photometry} and shown in Figure~\ref{fig:LC}. 

\subsection{Late optical epoch with GTC}
\label{sec:gtc}

EP260119a was observed with the High-speed Imaging Photometer for Occultations \cite[HIPERCAM;][]{Dhillon2021} mounted on the 10.4-m Gran Telescopio Canarias (GTC) (PI: Ag\"u\'i Fern\'andez). The observations were obtained on 2026 February 21 at 03:45:51.6 UTC (MJD 61092.15685) using the five simultaneous HIPERCAM filters $u_s$, $g_s$, $r_s$, $i_s$, and $z_s$. 

HIPERCAM provides simultaneous imaging in five optical bands, allowing a direct measurement of the spectral energy distribution without temporal interpolation between filters. The observations were performed using $2\times2$ pixel binning and were centred at RA, Dec = 10:32:27, +65:29:27 (J2000), fully covering the error box of EP260119a.

The observations were reduced using a self developed pipeline, which corrects for bias, flat field, does the image registration and combination of the individual frames into a median one using the \textsc{SWarp} package \citep{swarp}. The final stacked images have an exposure time of $45\times60$ s in each band. The astrometry was corrected using bright stars in the field. Our images were analysed with the COLIBRÍ ASU pipeline consistently to the COLIBRÍ/LT data, using nearby stars from the SDSS DR17 catalog \citep{SDSS17} for calibration.

\subsubsection{Other optical and infrared observations}

Further observations of the optical counterpart were carried out by different facilities, including LCO 1-m telescope \citep{43452}, MASTER \citep{43454}, SAO RAS \citep{43463}, Nordic Optical Telescope \citep{43464}, SVOM/VT \citep{43466}, Katzman Automatic Imaging Telescope \citep{43468}, GTC \citep{43469}, AZT-33IK 1.5~m telescope \citep{43476}, and Xinglong Observatory \citep{43496}. Table~\ref{tab:photometry} lists the available photometry from the literature together with the dataset presented in this work.

\subsection{Photometric and Spectroscopic Redshift}
\label{sec:redshift}

\begin{figure}
	\includegraphics[clip, width=\linewidth]{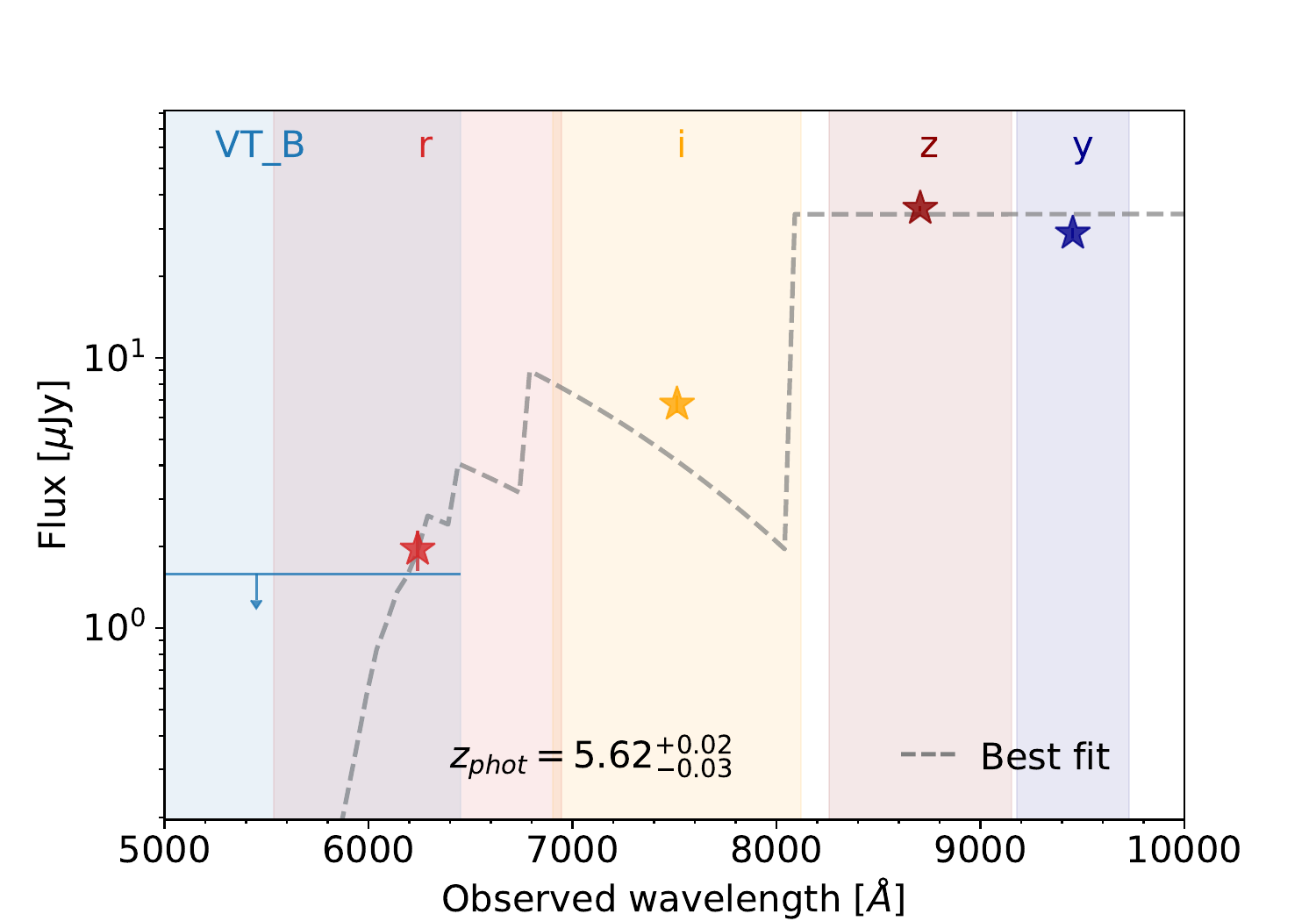}
    \caption{We show the optical SED obtained with COLIBRÍ in the $rizy$ bands at $T_0+12.65$~hours. We estimated the photometric redshift by modelling the distribution with the \texttt{zmodule} package. The dashed line indicates the best fit to the data (1$\sigma$) with a no-dust extinction law for a value $z_{\rm phot} = 5.62^{+0.02}_{-0.03}$. Shaded areas indicate the width of the COLIBRÍ filters. 
    \label{fig:photoz}}
\end{figure} 

We estimated the photometric redshift of the optical transient using temporally coincident \emph{rizy} images obtained with COLIBRÍ on January 19, 2026. The spectral energy distribution (SED) was modelled using the \texttt{zmodule}\footnote{\url{https://github.com/nyavorak/zmodule}} package \citep{Corre2018}. The fit was performed assuming negligible host-galaxy dust extinction, appropriate for the sharp spectral break observed between the $i$ and $z$ bands.

The resulting best-fit redshift is $z_{\rm phot} = 5.62^{+0.02}_{-0.03}$ $ (1\sigma, \mathrm{stat})$ driven by the pronounced Lyman-$\alpha$ break entering the observed optical bands (Figure~\ref{fig:photoz}). This measurement was first reported by \citet{43455}. 

Subsequent spectroscopic observations of the optical counterpart with the OSIRIS+ spectrograph mounted on the GTC \citep{Cepa1998} revealed a strong continuum break at $\sim 7870$~\AA, interpreted as the onset of intergalactic Lyman-$\alpha$ absorption. From this feature, \cite{43469} derived a spectroscopic redshift of $z = 5.47$.

The spectroscopic value differs from our photometric estimate by $\Delta z \approx 0.15$, i.e. $(z_{\rm spec} - z_{\rm phot})/(1+z_{\rm spec}) \approx 0.02$.
The uncertainty quoted above for $z_{\rm phot}$ is the statistical uncertainty of the SED fit only, not including systematic terms, constraining the redshift mainly by the fractional flux decrement in the band containing the break. The agreement between $z_{\rm spec}$ and $z_{\rm phot}$ at the $2\%$ level is good by this standard and confirms the robustness of the SED-based redshift determination and the high-$z$ nature of the transient.

With $z = 5.47$, EP260119a is currently the highest-redshift source detected by {\EP}, among both gamma-ray–associated and gamma-ray–quiet events.

\begin{table*}
    \centering
    \caption{Photometry of EP260119a used for this work. Values are not corrected by Galactic extinction. Upper limits are at $3\sigma$
    }
    \label{tab:photometry}
\begin{tabular}{lccccrr}
\toprule
\toprule
\multicolumn{7}{c}{\textbf{X-rays}}  \\
\toprule
\toprule
\textbf{Telescope} &\textbf{Instrument} & \textbf{Exposure} & \textbf{Mid Time} & \textbf{Energy} & \textbf{Flux} & \textbf{Reference}\\
& & \textbf{[s]} & \textbf{[h]} &\textbf{Range} &\textbf{[erg s$^{-1}$ cm$^{-2}$]} &  \\
\midrule
\midrule
EP&WXT              & \ldots &  0.042 & 0.5--4       & $(4.4^{+1.8}_{-1.3})\times 10^{-11}$ & \cite{43447} \\
EP&FXT              & 6000 &  8.33 & 0.5--10       & $(5.69^{+0.60}_{-0.55})\times 10^{-13}$ & \cite{43465} \\
\toprule
\toprule
\multicolumn{7}{c}{\textbf{Optical}}  \\
\toprule
\toprule
\textbf{Telescope}& \textbf{Instrument}  & \textbf{Exposure} & \textbf{Mid Time} & \textbf{Filter} & \textbf{AB Magnitude} & \textbf{Reference} \\
& & \textbf{[s]} & \textbf{[h]} & & &  \\
\midrule
\midrule
COLIBRÍ&DDRAGO      &  600 &  8.87 & \emph{z}     & 20.07 $\pm$ 0.03 & This work \\
COLIBRÍ&DDRAGO      & 1200 &  9.04 & \emph{r}     & 23.02 $\pm$ 0.12 & This work \\
COLIBRÍ&DDRAGO      &  600 &  9.12 & \emph{z}     & 20.08 $\pm$ 0.04 & This work \\
COLIBRÍ&DDRAGO      &  600 &  9.36 & \emph{z}     & 20.11 $\pm$ 0.04 & This work \\
COLIBRÍ&DDRAGO      & 1200 &  9.53 & \emph{r}     & 22.89 $\pm$ 0.10 & This work \\
LCO &   SINISTRO              &  900 &  9.54 & \emph{r}     & 22.83 $\pm$ 0.33 & \cite{43452} \\
COLIBRÍ&DDRAGO      &  600 &  9.56 & \emph{z}     & 20.05 $\pm$ 0.03 & This work \\
COLIBRÍ&DDRAGO      &  600 &  9.81 & \emph{z}     & 20.13 $\pm$ 0.03 & This work \\
LCO  & SINISTRO              &  900 &  9.81 & \emph{z}     & 20.26 $\pm$ 0.13 & \cite{43452} \\
COLIBRÍ&DDRAGO      & 1200 &  9.99 & \emph{r}     & 23.03 $\pm$ 0.11 & This work \\
COLIBRÍ&DDRAGO      &  600 & 10.06 & \emph{z}     & 20.16 $\pm$ 0.04 & This work \\
COLIBRÍ&DDRAGO      &  600 & 10.46 & \emph{z}     & 20.15 $\pm$ 0.04 & This work \\
SVOM&VT             & 3075 & 10.66 & \emph{VT\_R} & 20.86 $\pm$ 0.08 & \cite{43466} \\
SVOM&VT             & 3250 & 10.70 & \emph{VT\_B} &     $>$23.40     & \cite{43466} \\
COLIBRÍ&DDRAGO      &  600 & 10.87 & \emph{z}     & 20.12 $\pm$ 0.04 & This work \\
COLIBRÍ&DDRAGO      &  660 & 11.49 & \emph{i}     & 21.85 $\pm$ 0.08 & This work \\
COLIBRÍ&DDRAGO      & 1200 & 11.52 & \emph{r}     & 23.14 $\pm$ 0.12 & This work \\
COLIBRÍ&DDRAGO      &  660 & 11.72 & \emph{i}     & 21.83 $\pm$ 0.08 & This work \\
COLIBRÍ&DDRAGO      &  660 & 12.04 & \emph{i}     & 21.80 $\pm$ 0.08 & This work \\
COLIBRÍ&DDRAGO      & 1920 & 12.37 & \emph{y}     & 20.25 $\pm$ 0.07 & This work \\
COLIBRÍ&DDRAGO      &  660 & 12.55 & \emph{i}     & 21.79 $\pm$ 0.09 & This work \\
COLIBRÍ&DDRAGO      &  540 & 12.73 & \emph{r}     & 23.02 $\pm$ 0.21 & This work \\
COLIBRÍ&DDRAGO      &  600 & 12.47 & \emph{z}     & 20.28 $\pm$ 0.06 & This work \\
COLIBRÍ&DDRAGO      &  540 & 13.03 & \emph{i}     & 21.73 $\pm$ 0.16 & This work \\
COLIBRÍ&DDRAGO      &  360 & 13.06 & \emph{z}     & 20.26 $\pm$ 0.11 & This work \\
Mondy&AZT-33IK      & 5040 & 17.83 & \emph{I}     & 20.87 $\pm$ 0.11 & \cite{43476} \\
SAO RAS &            & 2400 & 22.61 & \emph{Ic}    & 20.80 $\pm$ 0.10 & \cite{43463} \\
NOT    & ALFOSC            & 1800 & 23.24 & \emph{z}     & 20.94 $\pm$ 0.08 & \cite{43464} \\
NOT     & ALFOSC           & 1800 & 23.80 & \emph{i}     & 21.85 $\pm$ 0.06 & \cite{43464} \\
Liverpool Telescope & IO:O& 1200 & 24.37 & \emph{r}     &     $>$22.64     & This work  \\
Liverpool Telescope & IO:O& 1200 & 24.58 & \emph{i}     & 22.04 $\pm$ 0.22 & This work \\
Liverpool Telescope & IO:O& 1100 & 24.93 & \emph{z}     & 21.00 $\pm$ 0.21 & This work \\
GTC&OSIRIS+         & 4800 & 26.80 & \emph{z}     & 21.06 $\pm$ 0.10 & \cite{43469} \\
COLIBRÍ&DDRAGO      & 7560 & 35.27 & \emph{i}     & 23.25 $\pm$ 0.12 & This work \\
COLIBRÍ&DDRAGO      & 7200 & 35.30 & \emph{z}     & 21.53 $\pm$ 0.05 & This work \\
AZT-20 &             & 3780 & 39.24 & \emph{i'}    & 22.70 $\pm$ 0.20 & \cite{43500} \\
Mondy&AZT-33IK      & 5400 & 43.39 & \emph{I}     &     $>$22.30     & \cite{43499} \\
SAO RAS&Zeiss-1000  & 2400 & 51.23 & \emph{Ic}    &     $>$21.90     & \cite{43499} \\
COLIBRÍ&DDRAGO      & 9360 & 58.30 & \emph{i}     & 24.00 $\pm$ 0.29 & This work \\
COLIBRÍ&DDRAGO      & 9240 & 58.28 & \emph{z}     & 22.68 $\pm$ 0.20 & This work \\
Mondy&AZT-33IK      & 6240 & 67.78 & \emph{I}     &     $>$21.70     & \cite{43499} \\
\bottomrule
\bottomrule

\multicolumn{7}{c}{\textbf{Candidate Host Galaxy}}  \\
\toprule
\toprule
\textbf{Telescope}& \textbf{Instrument}  & \textbf{Exposure} & \textbf{Mid Time} & \textbf{Filter} & \textbf{AB Magnitude} & \textbf{Reference} \\
& & \textbf{[s]} & \textbf{[d]} & & &  \\
\midrule
\midrule
GTC&HIPERCAM      &  2700 &  33.2 & $u_\mathrm{s}$     & $>25.67$& This work \\
GTC&HIPERCAM      &  2700 &   33.2  & $g_\mathrm{s}$     & $>27.19$ &This work \\
GTC&HIPERCAM      &  2700 &   33.2  & $r_\mathrm{s}$    & 25.52 $\pm$ 0.21 & This work \\
GTC&HIPERCAM      &  2700 &   33.2  & $i_\mathrm{s}$     & 25.10 $\pm$ 0.18 & This work \\
GTC&HIPERCAM      &  2700 &   33.2 & $z_\mathrm{s}$     & 25.18 $\pm$ 0.37 & This work \\
\bottomrule
\bottomrule

\end{tabular}
\end{table*}

\section{Absence of Gamma-Ray Emission and Classification of EP260119a as a GQFXT}
\label{sec:gammaray}

To establish whether EP260119a belongs to the emerging class of gamma-ray quiet fast X-ray transients (GQFXTs), we searched for contemporaneous prompt gamma-ray emission using all available high-energy instruments.

No prompt gamma-ray counterpart was detected in association with EP260119a. At the time of the EP/WXT detection, the {\itshape Fermi Space Telescope} \citep{Meegan2009} was within the South Atlantic Anomaly, preventing meaningful observations with both GBM and LAT. Nevertheless, the source was simultaneously covered by \textit{SVOM}, \textit{Swift}, and {\KW}, providing contemporaneous high-energy coverage despite the unavailability of \textit{Fermi}.

EP260119a remained within the field of view of {\itshape SVOM}/ECLAIRs \citep{ECLAIRs2014,Godet2026}, {\itshape SVOM}/GRM \citep{Dong2010,He2025}, and \textit{Swift}/BAT \citep{Gehrels2004,Barthelmy2005} throughout the EP/WXT detection.

Neither the onboard nor the ground-based \textit{SVOM}/ECLAIRs trigger software detected any significant emission. We derive a $3\sigma$ upper limit on the 4-120~keV flux of $1.0\times10^{-9}$~erg~s$^{-1}$~cm$^{-2}$ in the interval between $T_0$ and $T_0 + 300$~s, assuming a power-law spectrum with a photon index of 1.8 \citep{43447}. We plot this value in Fig.~\ref{fig:sedhe}.

EP260119a was also within the field of view of \textit{SVOM}/GRM \citep{Dong2010,He2025}, which recorded no trigger associated with the event. Using a 300~s integration time, we derive $3\sigma$ upper limits in the 15-5000~keV band for different assumed spectral models. For typical long-GRB spectral parameters ($\alpha=-1.9$, $\beta=-3.7$, and $E_{\rm peak}=70$~keV), the corresponding upper limit is $3.1\times10^{-9}$~erg~cm$^{-2}$~s$^{-1}$ (Figure~\ref{fig:sedhe}).

EP260119a was within the field of view of the \textit{Swift}/BAT instrument for nearly the entire duration of the X-ray emission reported by WXT. No significant gamma-ray emission was detected during this interval. In total, The BAT survey data consist of 300~s exposure images and are limited to an energy range of $\sim14$--$45$~keV, rather than the nominal $14$--$195$~keV band. From the 300~s survey exposure beginning at 2026-01-19T00:13:59~UT ($T_0+11$~s), the \textit{Swift}/BAT team derived a $14$–$45$~keV flux upper limit (3$\sigma$) of $\sim 2.22 \times 10^{-9}$~erg~cm$^{-2}$~s$^{-1}$, assuming a power-law spectrum. No on-board \textit{Swift}/BAT trigger was recorded at shorter timescales. Furthermore, inspection of the event-rate data (summed over all detectors, with time resolution down to 64~ms) revealed no evidence of burst-like emission.

{\KW} \citep{Aptekar1995} was also operating normally during the event and detected neither a triggered nor a waiting-mode burst. Analysis of the waiting-mode data revealed no statistically significant excess around the time of the EP/WXT detection. For a typical long-GRB spectrum, the corresponding 90\% confidence upper limit on the peak flux is
$1.3\times10^{-7}\,{\rm erg\,cm^{-2}\,s^{-1}}$ (see Figure~\ref{fig:sedhe}) in the 20--1500~keV band on a 2.944~s timescale.
The corresponding fluence upper limit is $6.3\times10^{-7}\,{\rm erg\,cm^{-2}}$ for the assumptions described above. The resulting upper limits are shown in Figure~\ref{fig:sedhe}.

Because the prompt gamma-ray spectrum is not constrained by the available observations, we construct an illustrative Band-function spectrum \citep{Band+93} that is consistent with the EP/WXT soft X-ray measurement and the contemporaneous gamma-ray upper limits (see Figure~\ref{fig:sedhe}). As an additional physical constraint, we assume that the event follows the Amati relation \citep{Amati+2002,Amati-06}, adopting
$E_{\rm peak}/{\rm keV} = 81\pm0.2 ( E_{\gamma, \rm iso}/10^{52} {\rm erg})^{0.57\pm0.02}$.  
The purpose of this exercise is not to uniquely reconstruct the prompt emission, but rather to identify a physically plausible prompt spectrum that simultaneously satisfies the available observational constraints and the assumed Amati relation.

We adopt fixed photon indices of $\alpha=-1.8$, matching the photon index measured by EP/WXT, and $\beta=-2.5$, representative of long GRBs \citep{Kaneko2006,Nava2011}. 
Following the procedure described in appendix \ref{appendix:band-amati}, this construction yields an isotropic-equivalent gamma-ray energy of $E_{\gamma,\rm iso}=(2.08^{+3.58}_{-1.32})\times10^{51}\,\mathrm{erg}$ computed over the standard 1-10,000~keV rest-frame energy interval. We stress that neither $E_{\rm peak}$ nor $E_{\gamma,\rm iso}$ are directly measured. These values represent one possible realization of the prompt emission under the adopted Band-function and Amati-relation assumptions. Other combinations of spectral parameters could also satisfy the EP/WXT measurement and the current gamma-ray upper limits.

As shown in Figure~\ref{fig:sedhe}, the illustrative spectrum remains below the upper limits derived from SVOM/ECLAIRs, \textit{SVOM}/GRM, \textit{Swift}/BAT, and {\KW}. The available gamma-ray observations therefore do not meaningfully constrain the location of the spectral peak or the detailed shape of the high-energy spectrum. Nevertheless, the independent contemporaneous non-detections by all these facilities show that no prompt gamma-ray counterpart was detected down to the reported instrumental limits. We therefore classify EP260119a observationally as a GQFXT. Whether its intrinsic gamma-ray emission was unusually weak cannot be determined without additional assumptions about the prompt spectrum.

Taken together, the independent contemporaneous non-detections by SVOM/ECLAIRs, SVOM/GRM, \textit{Swift}/BAT, and {\KW} show that no prompt gamma-ray counterpart was detected down to the reported instrumental limits. EP260119a can therefore be classified observationally as a GQFXT. 

\section{Environment}
\label{sec:hg}

\subsection{An extended source near the optical counterpart}

\begin{figure*}
	\includegraphics[clip, width=\linewidth]{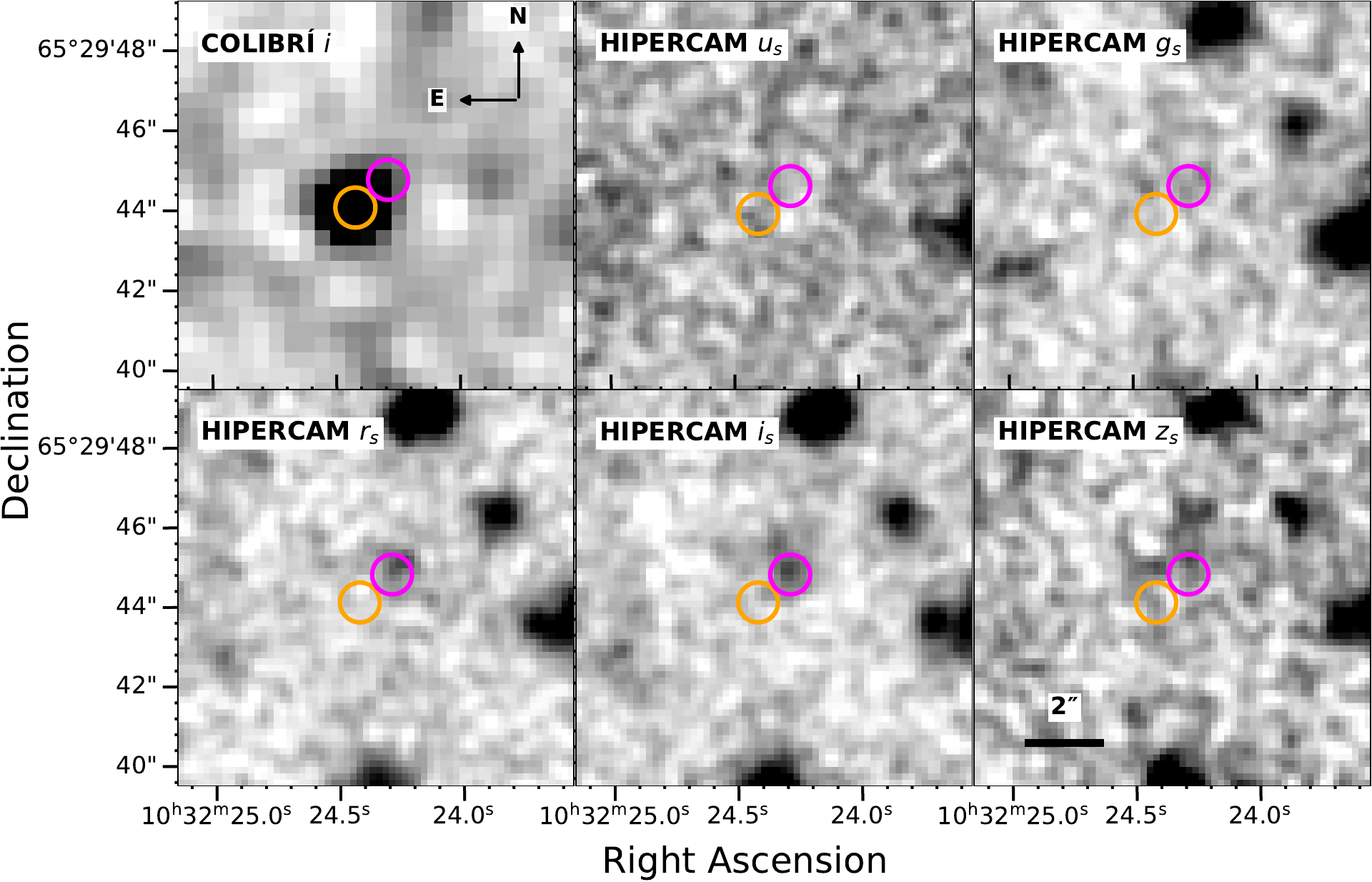}
    \caption{ $i$-band image compared with our late epoch acquired with the GTC/HIPERCAM instrument in the $u_s$, $g_s$, $r_s$, $i_s$, and $z_s$ filters of the field of EP260119a. The orange circle marks the position of the optical counterpart, while the magenta circle indicates the centroid of the candidate host galaxy. The scale bar in the lower-right panel corresponds to 2\arcsec.
    \label{fig:host}}
\end{figure*}

In our late images, we identify a faint extended source located at RA, DEC = 10:32:24.285, +65:29:44.790. The angular separation between the optical counterpart and the centroid of this extended source is $1.11 \pm 0.42$ arcsec. At a redshift of $z=5.47$, this corresponds to a projected physical offset of $6.57 \pm 4.34$ kpc (see Figure~\ref{fig:host}). Owing to its proximity, this source represents the most plausible host galaxy candidate detected in our images.

To estimate the photometry of the candidate host galaxy, we performed PSF photometry at the source position in the $u$, $g$, $r$, $i$, and $z$ images. We calibrated the photometry against the Sloan Digital Sky Survey (SDSS) catalog using stars brighter than 22 mag. The resulting host-galaxy magnitudes are $u > 25.7$ ($3\sigma$), $g > 27.2$ ($3\sigma$), $r = 25.50 \pm 0.21$, $i = 25.10 \pm 0.18$, and $z = 25.18 \pm 0.37$ mag (see Table~\ref{tab:photometry}). The red colours of the candidate are broadly consistent with expectations for a galaxy at high redshift, where absorption blueward of Ly$\alpha$ significantly suppresses the observed flux in the bluer filters.

\subsection{Photometric properties}

Unfortunately, owing to its faintness, obtaining a spectroscopic redshift for this candidate is not feasible. Nevertheless, we modelled its photometry after correcting for Galactic extinction with {\sc prospector} \citep{Johnson2021}, using \texttt{FSPS} stellar population models \citep{Conroy2009, Conroy2010}, adopting the \cite{Kroupa2001} initial mass function, a delayed-$\tau$ star-formation history with an age of $t_{\rm age}=0.5$~Gyr, an e-folding timescale of $\tau=0.3$~Gyr, and a fixed stellar metallicity of $Z=0.1~Z_\odot$. Intergalactic-medium absorption was included following \cite{Madau1995}, which suppresses the flux blueward of Ly$\alpha$ and determines the position of the Lyman break.

\begin{figure}
\includegraphics[clip,width=\linewidth]{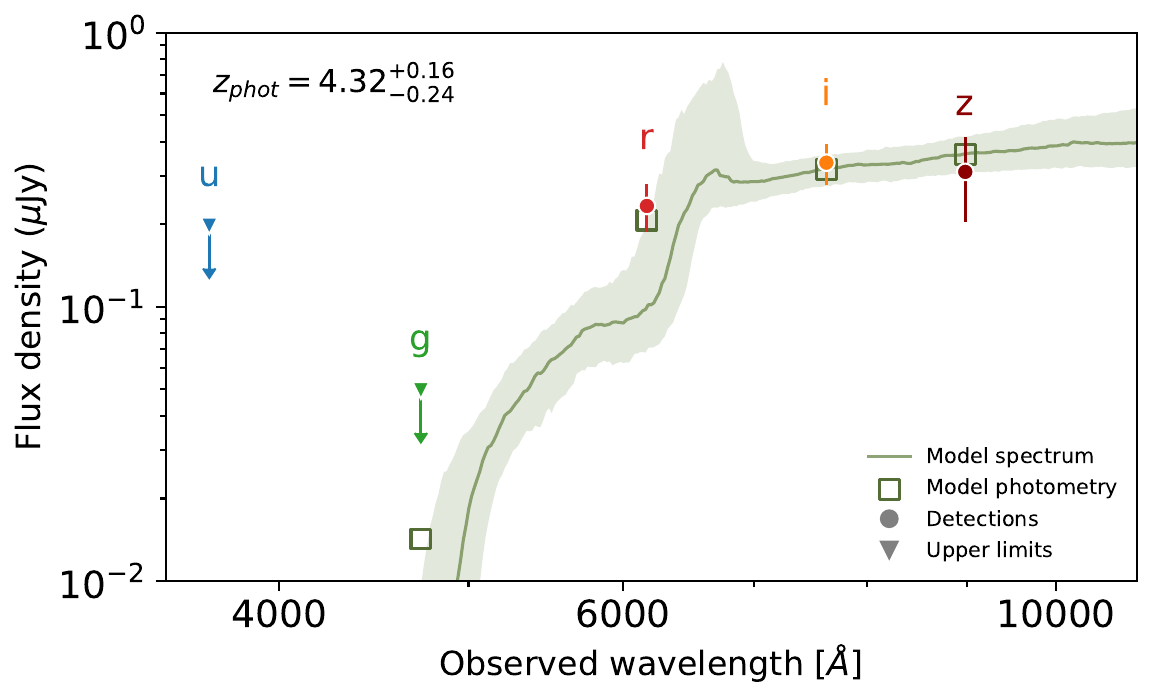} 
\caption{Spectral energy distribution of the candidate host galaxy of EP260119a. Coloured circles show the GTC/HIPERCAM detections in the $r$, $i$, and $z$ bands; downward triangles mark the $3\sigma$ upper limits in $u$ and $g$. The olive solid line and shaded band show the median and $1\sigma$ range of the posterior model spectrum from \texttt{Prospector} (described in Section~\ref{sec:hg}), with open squares giving the corresponding model photometry. Allowing the redshift to vary, the SED favours a photometric redshift of $z_{\rm phot} = 4.32 ^{+0.16}_{-0.24}$, set by the Lyman break falling between the $r$ and $i$ bands.}
\label{fig:sed_host}
\end{figure}

With the spectroscopic redshift of $z = 5.47$ (see Section~\ref{sec:redshift}), the model systematically underpredicts the observed $r$-band flux, indicating that the position of the Lyman break expected at this redshift is inconsistent with the measured colours. We therefore repeated the fit allowing the redshift, stellar mass, and dust attenuation $A_V$ to vary, and sampled the posterior distribution using nested sampling \citep[\texttt{dynesty};][]{Speagle2020}. Under these hypotheses, the SED fit instead favours a photometric redshift of $z_{\rm phot}=4.32^{+0.16}_{-0.24}$ (Figure~\ref{fig:sed_host}), which is not compatible with the spectroscopic redshift measured for the burst. 

The dust attenuation is only weakly constrained ($A_V \sim 0.7$) and remains degenerate with the stellar mass. In contrast, the scaling factor applied to the IGM opacity is consistent with the standard \cite{Madau1995} prescription ($f_{\rm IGM}=1.0\pm0.3$), indicating that the preferred redshift is not driven by intergalactic absorption. Nevertheless, the fit relies on only three significant photometric detections spanning the Lyman break, limiting the robustness of the inferred photometric redshift.

Although the observed photometry exhibits the characteristic suppression blueward of Ly$\alpha$ expected for a high-$z$ galaxy, the inferred position of the Lyman break is inconsistent with the spectroscopic redshift of EP260119a.
Taken together, the photometric-redshift discrepancy and the relatively large probability of chance coincidence ($P_\mathrm{cc}$; see the next subsection) suggest that the association between the detected galaxy and EP260119a remains uncertain. 

\subsection{Probability of chance coincidence}

An independent assessment of the association can be obtained from the probability of chance coincidence.

Using the extinction-corrected magnitude of the extended source ($r = 25.48$ mag), we estimate the probability of chance coincidence following the formalism of~\cite{Bloom2002} and the galaxy densities used by \citet{Becerra2023}. We obtain $P_{\rm cc}=0.07$ and a range of $P_{\rm cc}=0.06-0.09$ considering the photometric uncertainties. These values are significantly larger than the threshold typically adopted for robust host-galaxy associations, about $P_{\rm cc}<0.01$. Combined with the photometric redshift, which is incompatible with the spectroscopic redshift of EP260119a, these results strongly disfavour the detected galaxy as the host of EP260119a. The true host galaxy therefore remains undetected in our current observations.

\section{Temporal evolution of the light curve}
\label{sec:evolution}

\subsection{Afterglow Closure Relations}
\label{subsec:closure-relations}

Within the framework of the fireball model \citep{Rees1992, Meszaros1993, Meszaros1997}, the afterglow phase is interpreted as synchrotron emission produced by a population of shock accelerated electrons whose energy distribution follows a power law, $N(\gamma) \propto \gamma^{-p}$, where $\gamma$ is the electron Lorentz factor and $p$ is the power-law index. In this scenario, the observed flux density $F_\nu$ is typically described by a multi-segment power law in both time $T$ and frequency $\nu$, such that $F_{\nu} \propto T^{\alpha}\nu^{\beta}$ \citep{Sari1998,Granot2002}. 

We simultaneously model the light curve in the $i$, $r$, and $z$ bands with a smoothly broken power-law of the \cite{Beuermann1999} type, characterised by decay indices $\alpha_{\mathrm{early}}$ and $\alpha_{\mathrm{late}}$ for times $T < t_{\mathrm{break}}$ and $T > t_{\mathrm{break}}$, respectively, and smoothing parameter $s$:
\begin{equation}
F(T) = f_0 \left[\left(\frac{T}{t_{\mathrm{break}}}\right)^{-s \alpha_{\mathrm{early}}} +  \left(\frac{T}{t_{\mathrm{break}}}\right)^{-s \alpha_{\mathrm{late}}}\right]^{-1/s}.
\end{equation}
The break is assumed to be achromatic, so $\alpha_{\mathrm{early}}$, $\alpha_{\mathrm{late}}$, and $t_{\mathrm{break}}$ are treated as shared parameters across all three bands, while each band is assigned an independent normalisation $f_0$. To map the posterior parameter space and estimate uncertainties, we performed Markov Chain Monte Carlo (MCMC) analysis using the sampler \texttt{emcee} \citep{Foreman2013}. We adopted a Gaussian log-likelihood, assuming the photometric flux measurements in each band are independent, with normally distributed uncertainties given by their reported photometric errors, and applied uniform priors (See Appendix~\ref{app:mcmc-lc} for details) to the physical parameters to allow the data to drive the fit. The sampler was initialized with 64 walkers, discarded the first 2000 steps, and used the subsequent 20,000 steps for parameter estimation. The best fit parameters, representing the median of the distributions with uncertainties corresponding to the $1\sigma$ confidence intervals, are $\alpha_{\mathrm{early}} = -0.55^{+0.09}_{-0.08}$, $\alpha_{\mathrm{late}} = -2.24^{+0.21}_{-0.24}$, and $t_{\mathrm{break}} = 25.05^{+2.14}_{-2.13}$~h. Residuals ($F-F_{\rm{model}})/\sigma$ are shown in the bottom panel of Figure~\ref{fig:LC} and scatter within $2\sigma$ across all three bands. A single $i$-band measurement at $T\approx23.8$~h (NOT, \cite{43464}) deviates by $>5\sigma$ from the near contemporaneous Liverpool Telescope point. This point is retained in the fit for completeness and dominates the overall $\chi^2_{\nu} = 2.62$, while excluding it yields a more consistent $\chi^2_{\nu} = 0.91$. The physical parameters derived in Section~\ref{subsec:physical-fit} from a forward-shock model are unaffected by this choice.

The early decay, $\alpha_{\mathrm{early}}$, is shallower than expected for a standard forward shock in the slow-cooling regime, suggesting either continued energy injection or that the early observations do not yet sample the self-similar power-law evolution. Because these pre-break data span less than a decade in observed time, however, the inferred slope should be interpreted with caution.

After the initial shallow phase  ($T_0+25.03^{+2.14}_{-2.13}$~h), the optical emission of EP260119a seems to transition into a steep power-law decline with a temporal index of $\alpha_{\mathrm{late}} = -2.24^{+0.21}_{-0.24}$.
If the emission arises from a standard forward shock in the slow cooling regime ($\nu_m < \nu < \nu_c$), the temporal index is related to the electron energy distribution through $\alpha = 3(1-p)/4$ for a nearly  constant-density (ISM) medium and $\alpha = (1-3p)/4$ for a wind-like medium ($\rho \propto r^{-2}$) \citep{Sari1998,Granot2002}, which would require an unusually steep electron index, $p = 3.99^{+0.32}_{-0.28}$ (ISM) or $p = 3.32 ^{+0.32}_{-0.28}$ (wind), well outside the range $2 < p < 3$ typically inferred for GRB afterglows \citep[e.g.,][]{Kumar2015}.

Alternatively, if instead the steepening marks the onset of a jet break \citep{Zhang2004}, the post jet break relation for a sideways-expanding jet $\alpha \approx -p$, is largely independent of the circumburst density profile \citep{Granot2002} and yields  $p=2.24_{-0.21}^{+0.24}$, fully consistent with the canonical range. Within the standard synchrotron afterglow framework, the temporal evolution is more naturally explained by a post-jet-break scenario. Nevertheless, the limited post-break temporal sampling means this interpretation should be regarded as suggestive rather than definitive.

\subsection{MCMC Fits using a Uniform Jet}
\label{subsec:physical-fit}
To further investigate the broadband properties of the afterglow, we perform a full broadband fit using a physical forward-shock model.

Within the fireball model \citep{Rees1992, Meszaros1993, Meszaros1997}, we describe the broadband afterglow assuming an ultra-relativistic thin shell propagating into an external medium with a power-law density profile, $n(R)=A R^{-k}$, where $A = n_0 R_0^k$ ($n_0$ being the circumburst medium density at $R_0 = 10^{18}\, \rm{cm}$). The main stages of the dynamical evolution are summarized below.

(i). Initially, the shell coasts with a bulk Lorentz factor $\Gamma_0\gg1$ until it begins to decelerate by sweeping up the external medium. This interaction produces a pair of collisionless shocks: a forward shock propagating into the ambient medium and a reverse shock propagating back into the ejecta. The forward shock heats the swept-up material, which moves with a bulk Lorentz factor $\Gamma$ while the reverse shock extracts kinetic energy from the ejecta while shock-heating it.

(ii). Deceleration begins once the swept-up mass, $M_{\rm SW}(R)$, exceeds $E_{\rm k,iso}/(\Gamma_0^2c^2)$, where $E_{\rm k,iso}$ is the isotropic-equivalent kinetic energy of the ejecta. For a density profile $n(R)=A R^{-k}$, the swept-up mass is $M_{\rm SW}(R)=\int_0^R4\pi r^2m_pn(r)\,dr=4\pi m_pAR^{3-k}/(3-k)$ for $k<3$, yielding a deceleration radius $R_{\rm dec}=[(3-k)E_{\rm k,iso}/(4\pi m_pA\Gamma_0^2c^2)]^{1/(3-k)}$ and the corresponding observer-frame deceleration time $T_{\rm dec}\simeq(1+z)R_{\rm dec}/(2\Gamma_0^2c)$.

(iii). Beyond $R_{\rm dec}$, the forward shock follows the Blandford-McKee self-similar solution \citep{Blandford1976}, with $\Gamma \propto R^{-(3-k)/2}$.

During this phase, most of the shell kinetic energy is transferred to the shocked external medium. As the blast wave continues to decelerate, a jet break in the afterglow light curve occurs once $\Gamma(R)\simeq1/\theta_{\rm jet}$, where $\theta_{\rm jet}$ is the jet half-opening angle. This defines the jet-break radius $R_{\rm jb}=[(3-k)E_{\rm k,iso}/(4\pi m_pA\theta_{\rm jet}^{-2}c^2)]^{1/(3-k)}$ and the corresponding observer-frame jet-break time $T_{\rm jb}\simeq(1+z)R_{\rm jb}\theta_{\rm jet}^2/2c$.

The observed afterglow emission is produced by synchrotron radiation from electrons accelerated at the forward shock. A fraction of the electrons \citep[$m_{\rm e}/m_{\rm p}\leq \xi_e \leq 1$;][]{Eichler2005} 

entering the forward shock are accelerated into a power-law energy distribution, such that their comoving number density is $n_e(\gamma)=dn_e/d\gamma \propto \gamma^{-p}$ for $\gamma \geq \gamma_m$, where $\gamma$ is the electron Lorentz factor and $\gamma_m$ corresponds to the minimum electron energy. These electrons carry a fraction $\varepsilon_e$ of the post-shock internal energy. 
The collisionless shock also amplifies the upstream pre-existing magnetic field, which accounts for a fraction $\varepsilon_B$ of the downstream internal energy. The resulting synchrotron emission is described by a smoothly broken power-law spectrum with characteristic frequencies $\nu_m$ and $\nu_c$ \citep{Sari1998,Granot2002}. 
At late times, the afterglow is expected to be in the slow-cooling regime ($\nu_m<\nu_c$), where $\nu_m$ is the characteristic synchrotron frequency of electrons with Lorentz factor $\gamma_m$ and $\nu_c$ corresponds to electrons cooling on the dynamical timescale. In this regime, the spectral index of the flux density, $F_\nu\propto T^\alpha\nu^\beta$, is $\beta=1/3$ for $\nu<\nu_m$, $\beta=(1-p)/2$ for $\nu_m<\nu<\nu_c$, and $\beta=-p/2$ for $\nu>\nu_c$, where $T$ is the observer-frame time. Although the reverse shock can dominate the optical emission during the first $\sim10^3\,{\rm s}$ in some bursts \citep{Gao2015,Resmi2016,Sari1999,Kobayashi2000}, the forward shock dominates at later times. Thus, our modelling focuses exclusively on the forward-shock emission. 

We perform a light curve fit to the X-ray ($\nu_X=$ 1~keV) and optical ($\nu_O=3.44\times 10^{14}$~Hz) data using MCMC methods and a model of a top-hat jet \citep{Gill2018}, which is a uniform jet with sharp edges, at a polar angle $\theta_{\rm jet}$ measured from the jet symmetry axis, propagating in an external medium. We adopt the infinitely thin-shell approximation (i.e. an ejecta with no radial width), and we assume an on-axis observer ($\theta_{\rm obs} = 0$) and $\xi_e=1$. We fix the dust extinction to $A_V = 0.0272\, {\rm mag}$. 

\begin{figure}
    \centering
    \includegraphics[width=1.0\linewidth]{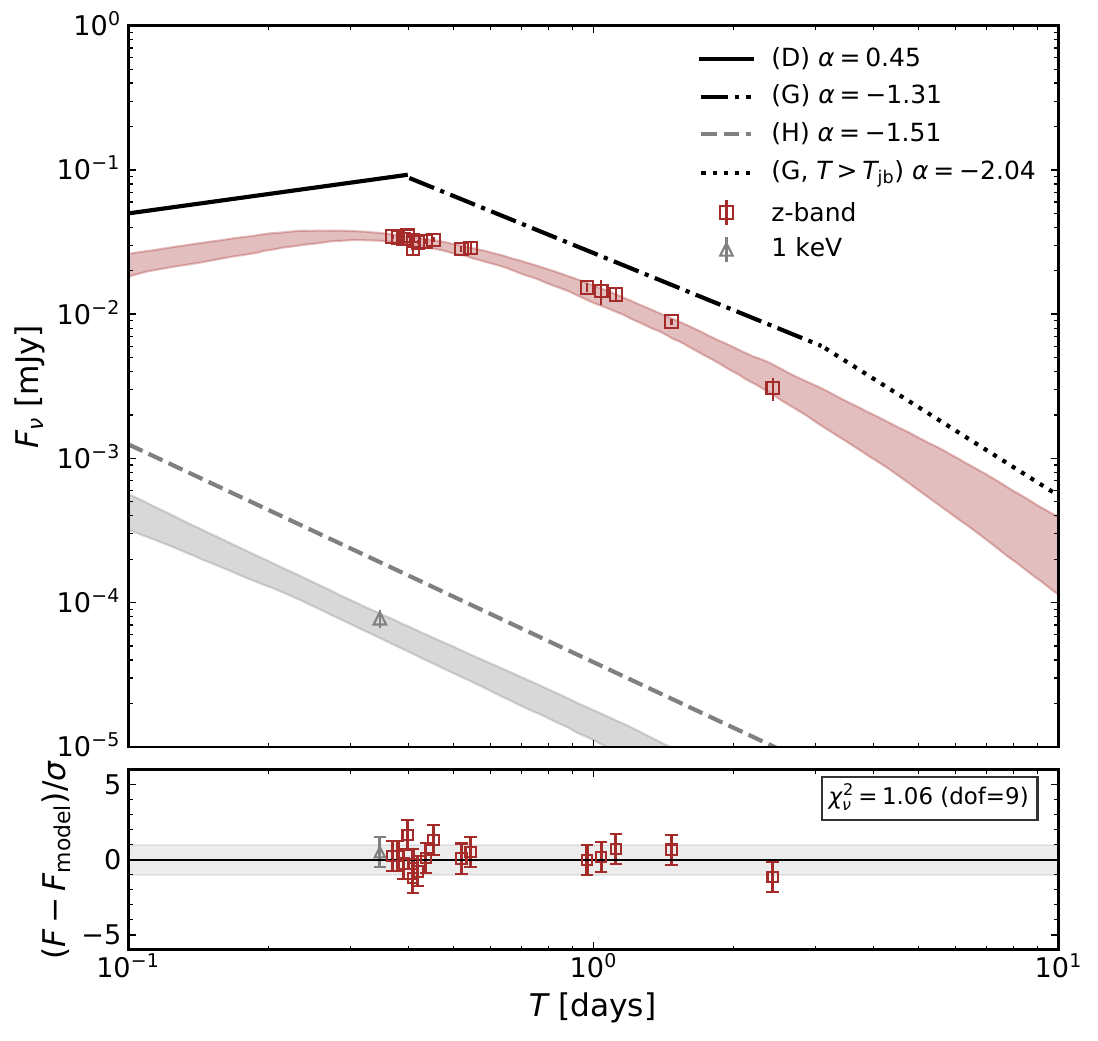}
    \caption{(Top:) Afterglow model light curve fit to observations using the model from \citet{Gill2018} that features a blast wave from a uniform jet propagating inside a radially stratified external medium. The brown (z-band) and grey (1\,keV) bands are the 1$\sigma$ uncertainty regions of the light curves obtained by randomly sampling the posterior distribution of model parameters. The solid, dashed-dotted, dashed, and dotted lines show the temporal power-laws calculated from \citet{Beniamini2020}, using the best fit parameters from MCMC analysis. (Bottom:) Residuals, $(F-F_{\rm model})/\sigma$, evaluated at the posterior-median model for each data point. The shaded band indicates the $\pm1\sigma$ region. The reduced $\chi^2$ calculated from z-band and X-ray data  ($\chi^2_\nu=1.06$) and its degrees of freedom (${\rm dof}=9$) are also shown.}
    \label{fig:lightcurve_top_hat_jet}
\end{figure}

Table~\ref{tab:MCMCparameters} lists the best-fit parameters and their 1$\sigma$ uncertainties obtained from the MCMC analysis. The corresponding light curves, MCMC prior ranges, and posterior distributions are shown in Figures~\ref{fig:lightcurve_top_hat_jet} (see the top panel), Table~\ref{tab:mcmc_priors_uniform_jet}, and Figure~\ref{fig:corner_plots_top_hat_jet}, respectively. A broad search for the best-fit parameters was initially performed using the wide priors shown in Table~\ref{tab:mcmc_priors_uniform_jet}. Narrow priors were used later to improve the sampling around the potential solution and obtain the posterior distribution of the model parameters in Figure~\ref{fig:corner_plots_top_hat_jet}. All inferred parameters lie within the typical range for GRB afterglows \citep{Ghisellini2009,Santana2014}, except for the relatively low circumburst density $n_0$. We show the residuals $(F-F_{\rm model})/\sigma$ in the bottom panel of Figure~\ref{fig:lightcurve_top_hat_jet}. The reduced $\chi^2$ calculated from the z-band and X-ray data is $\chi^2_\nu=1.06$ and its degrees of freedom (dof) is $9$. For comparison with the smoothly broken power-law fit in Section~\ref{subsec:closure-relations}, we also calculate $\chi^2_\nu$ using only the z-band data, yielding $\chi^2_\nu=1.16$ with dof $=8$.

The best-fit MCMC value of $E_{\rm k,iso}$ and its 1$\sigma$ uncertainty implies a prompt $\gamma$-ray efficiency of $\eta_\gamma=[1+(E_{\rm k,iso}/E_{\gamma,\rm iso})]^{-1}=2.61^{+6.92}_{-1.87}$ per cent. The central value is significantly lower than the values typically inferred ($\eta_\gamma\sim10-20$ per cent) for GRBs \citep{Beniamini2016}, however, the propagated upper uncertainty reaches $\sim 9.5$ per cent, approaching to the lower end of this typical range. This estimate should nevertheless be interpreted with caution,  as it is conditional on the assumed prompt spectrum, since $E_{\gamma,\rm iso}$ is not directly measured and is instead inferred from the illustrative Band-function spectrum constrained through the Amati relation. 
Using the best-fit values of $\Gamma_0$, $E_{\rm k,iso}$, $n_0$, and $k$, together with their 1$\sigma$ uncertainty, we derive an upper limit on the deceleration time of $T_{\rm dec}\simeq(1+z)R_{\rm dec}/(2\Gamma_0^2c)<5.5\times10^2\,{\rm s}$. 
Owing to the lack of afterglow observations before 0.3 days, the uncertainty remains large, and only an upper limit can be derived. Similarly, we estimate a jet-break time of $T_{\rm jb}\simeq(1+z)R_{\rm jb}/(2\theta_{\rm jet}^{-2}c)=3.1\pm2.0\,{\rm days}$. The large uncertainty reflects the absence of observations beyond $\sim3\,{\rm days}$.

\begin{figure}
    \centering
    \includegraphics[width=1.0\linewidth]{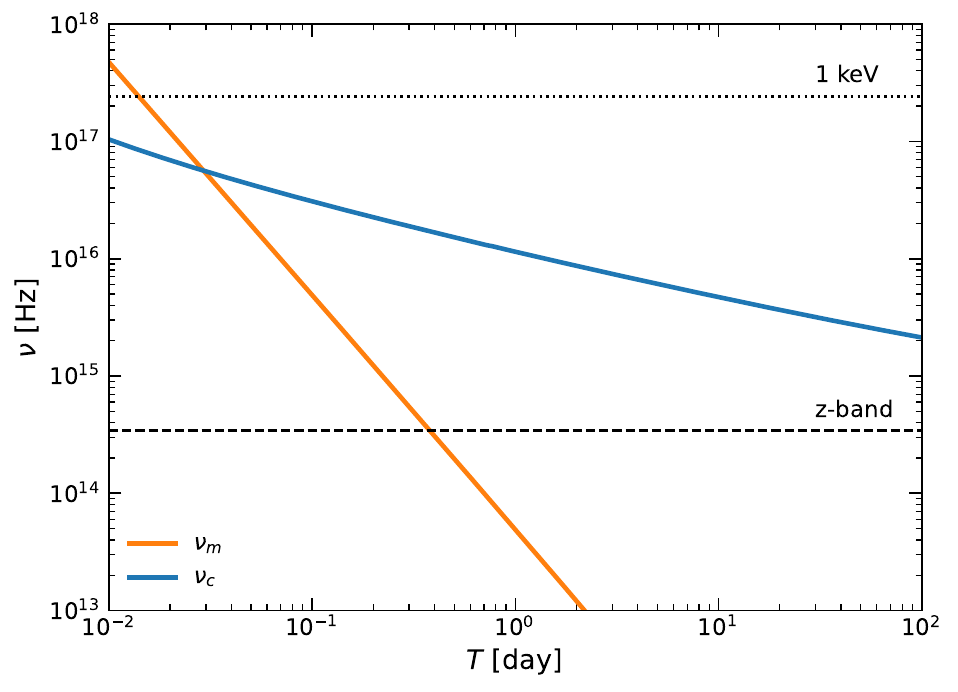}
    \caption{Temporal evolution of $\nu_m$ (orange line) and $\nu_c$ (blue line).}
    \label{fig:numnuc}
\end{figure}

From the best-fit parameters, we derive the temporal evolution of $\nu_m$ and $\nu_c$ (Figure~\ref{fig:numnuc}). Throughout the X-ray observations, the condition $\nu_X>{\rm max}(\nu_m,\nu_c)$ is satisfied, resulting in a single power-law decay with $\alpha=(2-3p)/4\simeq-1.5$ for $T>0.02\,{\rm days}$. 
The optical afterglow initially rises with $\alpha=(2-k)/(4-k)\simeq0.45$, reaching its peak at $T\sim0.4\,{\rm days}$ while $\nu_O<\nu_m<\nu_c$ for $0.03<T<0.4\,{\rm days}$. Once $\nu_m$ crosses the optical band, the light curve enters the $\nu_m<\nu_O<\nu_c$ regime and decays with $\alpha=[k(3p-5)-12(p-1)]/4(4-k)\simeq-1.3$. Finally, the jet break at $T_{\rm jb}\sim 3\,{\rm days}$ steepens the decay to $\alpha=(k+12p-3kp)/(4(k-4))\simeq-2.0$ \citep{Beniamini2020}. 

The EP/FXT observation gives an X-ray photon index of $\Gamma_X=1.99\pm 0.13$ (see Section~\ref{sec:xray-obs}). For $\nu_X>{\rm max}(\nu_m,\nu_c)$, as suggested by the above fitting results, this corresponds to $p=1.98\pm0.26$, which is somewhat smaller than the value obtained from our fit. If the X-rays is in the regime $\nu_m < \nu_X < \nu_c$, the corresponding value is $p=2.98\pm0.26$, which is closer to the value from our fit. However, the EP/FXT observation is obtained at only a single epoch, and we do not have information on the temporal evolution of the X-ray flux. The X-ray light curve is needed to determine the spectral regime and constrain the value of $p$ more accurately.

The temporal break at $T \sim 1\,{\rm day}$ seen in Figure~\ref{fig:LC} could be interpreted as the jet break, as discussed in Section~\ref{subsec:closure-relations}, since $T_{\rm jb}$ has a $1\sigma$ range of $1.1 \leq T_{\rm jb} \leq 5.1\,{\rm days}$, which includes $\sim 1\,{\rm day}$. We emphasize, however, that the phenomenological break time $t_{\rm break}$ obtained from the smoothly broken power-law fit and the physical jet-break time $T_{\rm jb}$ inferred from the broadband model are conceptually distinct quantities and need not correspond one-to-one.

Alternatively, this observing steepening may not require a distinct physical emission component. In the broadband afterglow model, it arises naturally from the smooth evolution of the synchrotron spectrum combined with equal-arrival-time surface effects \citep{Granot1999}, which broaden the transition and produce an extended shallow phase before the subsequent decay.

We also examine the possibility that the afterglow is produced by an off-axis top-hat jet, which is one of the proposed explanations for FXTs \cite{Sun2025,Ricci2025,Li2026}. 
To investigate this model, we treat $\theta_{\rm obs}$ as a free parameter and perform an MCMC analysis. The best-fit parameters correspond to an on-axis configuration, i.e., $\theta_{\rm obs} < \theta_{\rm jet}$. We also perform an MCMC analysis with the prior constraint $\theta_{\rm obs} > \theta_{\rm jet}$, but find no acceptable solution. Therefore, the off-axis top-hat jet model does not provide a satisfactory fit within the explored parameter space and assumptions.

In the above model, we assume $\xi_e=1$. In general, there is a degeneracy associated with $\xi_e$. The parameters obtained under the assumption $\xi_e=1$ is denoted as $(E_{\rm k,iso}^*, n_0^*, \varepsilon_e^*, \varepsilon_B^*)$. For any $m_{\rm e}/m_{\rm p} \leq \xi_e \leq 1$, the parameter sets $(E_{\rm k,iso}, n_0, \varepsilon_e, \varepsilon_B) = (\xi_e^{-1} E_{\rm k,iso}^*, \xi_e^{-1} n_0^*, \xi_e \varepsilon_e^*, \xi_e \varepsilon_B^*)$ provide equally good fits to the data \citep{Eichler2005}. For example, if we treat $\xi_e$ as a free parameter, $\gamma_m \propto \varepsilon_e/\xi_e$ and $\nu_m \propto (\varepsilon_e/\xi_e)^2$. From our fitting results, $\nu_m$ crosses the optical band at the peak time of the optical flux. In order to keep $\nu_m$ unchanged at the crossing time, the value of $\varepsilon_e$ depends on the choice of $\xi_e$. This degeneracy may be broken by accounting for the emission, absorption, or propagation effects of thermal electrons \citep{Sagiv2004, Toma2008, Giannios2009, Ressler2017, Warren2022}. However, given the sparse data, including the lack of radio observations and the availability of only a single X-ray data point, it is difficult to constrain $\xi_e$ for the EP 260119a afterglow. Introducing $\xi_e$ as an additional free parameter would not provide a meaningful improvement in the fit. Thus, we fix $\xi_e=1$ in our fitting.

\begin{table*}
\centering
\caption{Model parameters from the MCMC analysis.}
\label{tab:MCMCparameters}
\begin{tabular}{cccccccc}
\hline
$\log_{10} (\theta_{\rm jet}/{\rm rad})$ & $\log_{10} (E_{k,\rm iso}/{\rm erg})$ & $\log_{10} \Gamma_{\rm 0}$ & $\log_{10} (n_0/{\rm cm^{-3}})$ & $p$ & $\log_{10} \varepsilon_e$ & $\log_{10} \varepsilon_B$ & $k$ \\
\hline
$-1.33^{+0.22}_{-0.11}$ & $52.89^{+0.12}_{-0.16}$ & $2.46^{+0.33}_{-0.40}$ & $-1.79^{+0.18}_{-0.13}$ & $2.68^{+0.18}_{-0.22}$ & $-0.44^{+0.03}_{-0.04}$ & $-2.35^{+0.39}_{-0.29}$ & $0.39^{+0.33}_{-0.25}$\\
\hline
\end{tabular}
\end{table*}

\section{Discussion}
\label{sec:discussion}

\subsection{Luminosity and Energy}

\begin{figure}
	\includegraphics[clip, width=0.95\linewidth]{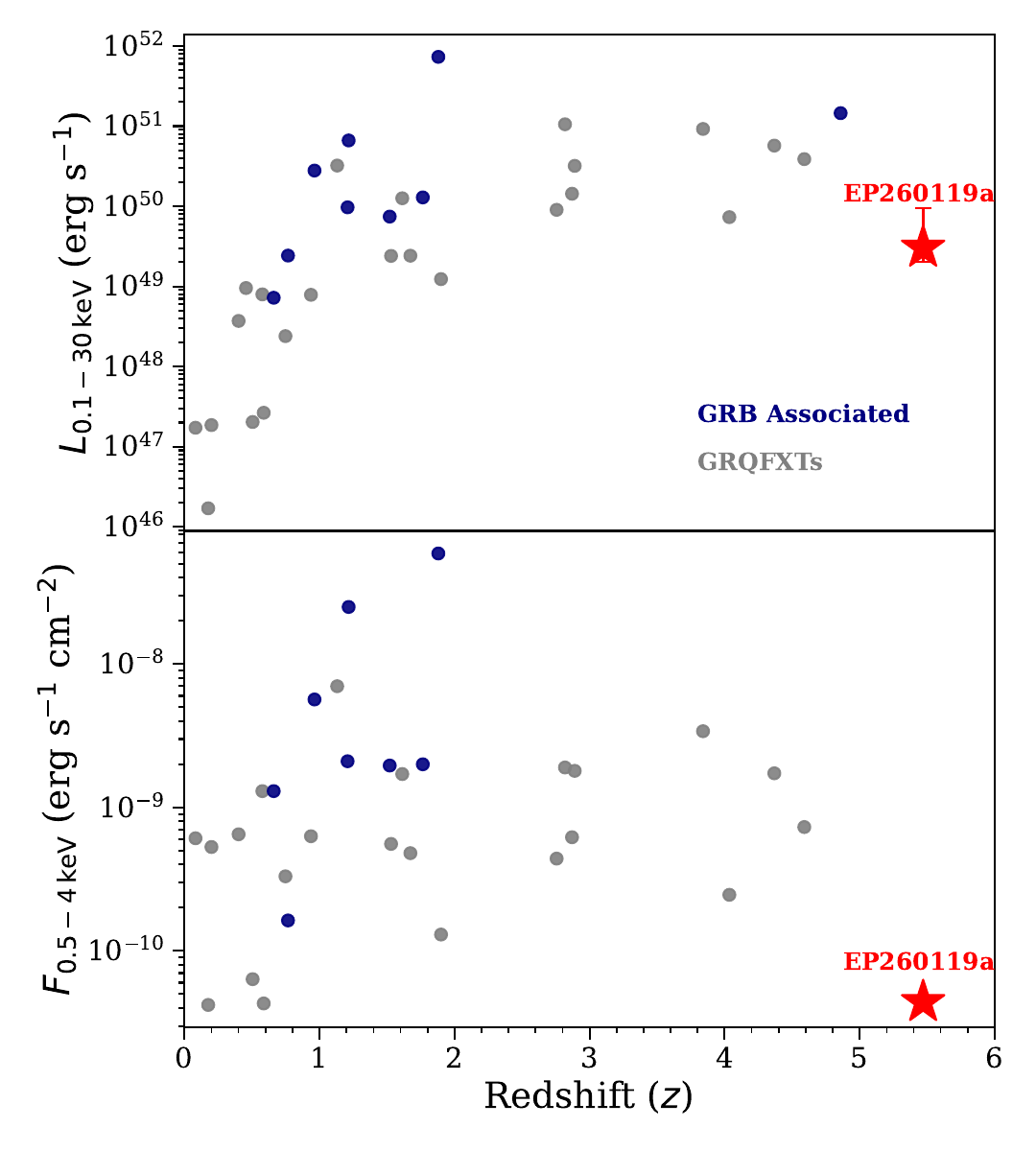}
    \caption{X-ray fluxes and luminosities of the {\EP} FXTs. Top: Isotropic-equivalent luminosities computed in a common rest-frame $0.1$--$30$~keV band assuming a power-law spectrum for each event.
    Bottom: Unabsorbed $0.5$--$4$~keV fluxes reported by \textit{EP}/WXT for all FXTs with a measured redshift.
    In both panels, FXTs associated with gamma-ray bursts are shown in blue, while those without an associated gamma-ray counterpart are shown in grey. EP260119a is highlighted with a red star.}
\label{fig:lumx}
\end{figure}

We present the average unabsorbed \textit{EP}/WXT $0.5$-$4$~keV flux as a function of redshift, without applying any $k$-correction or band pass conversion. In the top panel of Figure~\ref{fig:lumx}, we illustrate that the fluxes correspond directly to the measured instrumental energy range and therefore reflect both the intrinsic luminosity and the effects of cosmological dimming. This provides a model-independent view of the observed distribution and allows EP260119a to be assessed relative to the raw FXT population. 

However, we have to consider the cosmological effects to make a clean comparison. At a redshift of $z=5.47$ ($D_L \simeq 53$~Gpc) (see Section~\ref{sec:redshift}), corresponding to $(1+z)=6.47$, the observed photon energies are shifted to higher energies in the source rest frame according to
\begin{equation}
E_{\rm rest} = (1+z)\,E_{\rm obs}.
\end{equation}

Therefore, the observed \textit{EP}/WXT $0.5-4$~keV band corresponds to a rest-frame energy range of $3.24-25.9$~keV. The measured flux thus probes the hard X-ray regime in the burst rest frame and is less sensitive to soft X-ray absorption than it would be at lower redshift. However, direct comparisons with lower-$z$ events observed in the same instrumental band require a $k$-correction, since the corresponding rest-frame energy coverage differs.

At $z=5.47$ ($D_L\simeq53$~Gpc), the observed $EP$/WXT 0.5-4~keV band corresponds directly to the rest-frame 3.24-25.9~keV energy interval. Therefore, the isotropic-equivalent luminosity in this matched rest-frame band is given by $L_{3.24-25.9}=4\pi D_L^2 F_{0.5-4}$, without requiring an additional spectral $k$-correction. Using the average unabsorbed flux $F_{0.5-4}=4.4^{+1.8}{-1.3}\times10^{-11}\ \rm erg\ s^{-1}\ cm^{-2}$ \citep{43447}, we obtain an isotropic-equivalent luminosity of $L_{3.24-25.9}\simeq1.5\times10^{49}\ \rm erg\ s^{-1}$. The corresponding isotropic-equivalent peak luminosity is therefore $L^{\rm pk}_{3.24-25.9}\simeq1.6\times10^{50}\ \rm erg\ s^{-1}$, derived from the peak flux $F^{\rm pk}_{0.5-4}=4.8\times10^{-10}\ \rm erg\ s^{-1}\ cm^{-2}$.

To enable a homogeneous comparison between EP260119a and the broader FXT population, we restricted the sample to events with measured redshifts. We computed isotropic-equivalent luminosities in a common rest-frame $0.1$--$30$~keV band, following a standard definition of a fixed intrinsic energy interval which is sufficiently narrow for a power-law approximation while encompassing the soft X-ray emission of EP FXTs. Assuming a power-law spectrum $N(E)\propto E^{-\Gamma}$, the observed \textit{EP}/WXT fluxes were extrapolated from the observed $0.5$-$4$~keV band to the corresponding observer-frame interval $[0.1/(1+z),\,30/(1+z)]$~keV. The extrapolated flux was calculated as

\begin{equation}
F(E_a,E_b)=F_{0.5-4}
\frac{\displaystyle\int_{E_a}^{E_b}E^{1-\Gamma},dE}
{\displaystyle\int_{0.5}^{4}E^{1-\Gamma},dE},
\end{equation}
where $E_a=0.1/(1+z)$ and $E_b=30/(1+z)$. The corresponding isotropic-equivalent luminosity was then computed as

\begin{equation}
L_{0.1-30}=4\pi D_L^2
F\left(\frac{0.1}{1+z},\frac{30}{1+z}\right).
\end{equation}

For EP260119a, $z=5.47$ corresponds to an observer-frame interval of $0.0155$--$4.64$~keV. Using the best-fit photon index $\Gamma=1.8$ and the average unabsorbed flux $F_{0.5-4}=4.4\times10^{-11}\ \rm erg\ s^{-1}\ cm^{-2}$, we obtain $L_{0.1-30}\simeq3.05\times10^{49}\ \rm erg\ s^{-1}$. The relatively large uncertainty in the photon index, $\Gamma=1.8\pm0.6$, introduces a substantial and asymmetric uncertainty because the calculation requires extrapolation below the observed WXT band. Considering the photon-index uncertainty alone, we obtain $L_{0.1-30}=3.05^{+6.40}_{-1.02}\times10^{49}\ \rm erg\ s^{-1}$. This uncertainty is taken into account when comparing EP260119a with the broader FXT population.

This procedure ensures that all events are compared over the same intrinsic energy interval, thereby minimizing redshift-dependent bandpass effects. We show this comparison in Figure~\ref{fig:lumx}; the sample is divided according to the presence or absence of an associated gamma-ray counterpart.

Although EP260119a is the highest-redshift FXT currently known, its rest-frame X-ray luminosity lies within the range occupied by previously identified EP FXTs. This suggests that its observational uniqueness is primarily due to its distance rather than an exceptionally energetic prompt X-ray emission. Moreover, the currently known GQFXTs span a broad luminosity range comparable to GRB-associated FXTs, suggesting that the absence of detected gamma rays is not simply a consequence of intrinsically weak explosions.
A systematic comparison of the growing EP sample will be possible in a future work.

\subsection{Redshift distribution of GQFXTs}

\begin{figure}
	\includegraphics[width=\linewidth]{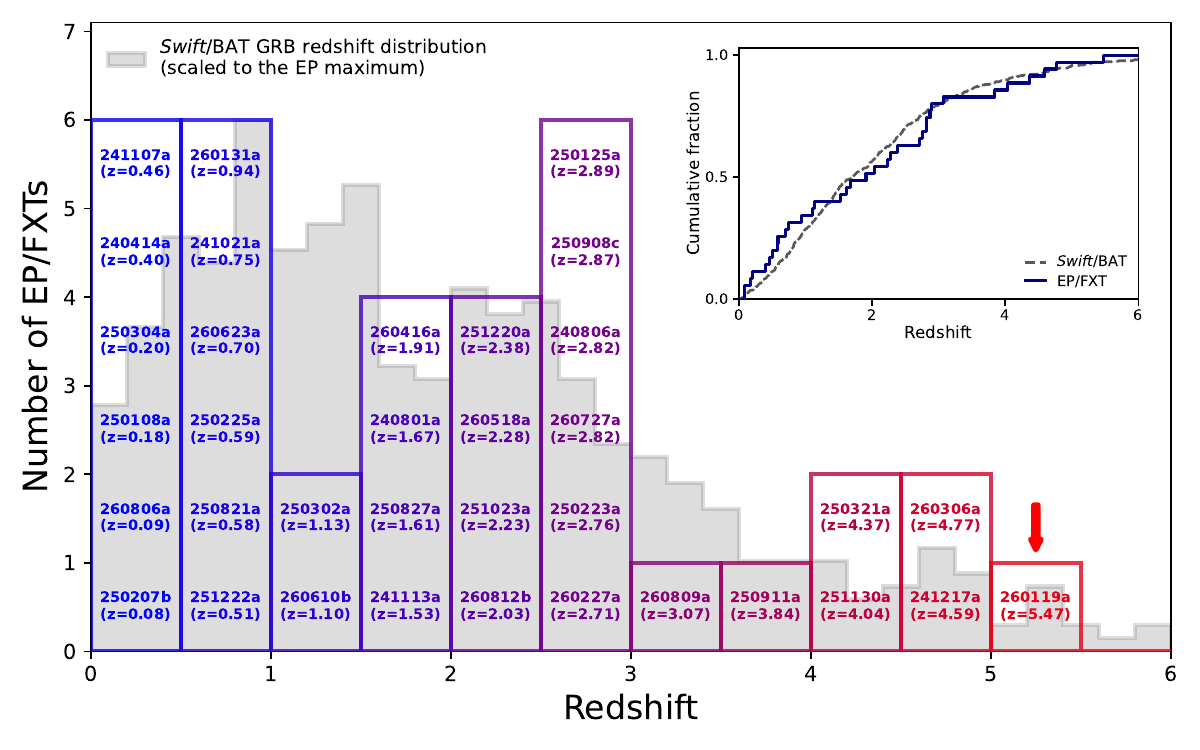}
    \caption{Redshift distribution of \textit{Einstein Probe} gamma-ray quiet fast X-ray transient with spectroscopic or photometric redshift measurements up until 4 September 2026. EP260119a is currently the highest-redshift member of the sample. Individual events are labelled within each bin. The shaded grey histogram shows the redshift distribution of \textit{Swift}/BAT GRBs, scaled such that its peak matches that of the \textit{Einstein Probe}/FXT distribution for comparison. The inset panel illustrates the cumulative distribution function (CDF) for the redshift distribution of GQFXTs (blue solid line) and the \textit{Swift}/BAT GRBs (dashed grey line). \label{fig:zhist}}
\end{figure}

EP260119a lies at the extreme high-$z$ end of the currently known {\EP} gamma-ray quiet fast X-ray transient population. As recently shown by \cite{OConnor2025}, the cumulative redshift distribution of \textit{Einstein Probe} FXTs closely follows that of long GRBs, suggesting that the two populations trace similar cosmic star-formation histories despite differences in their prompt high-energy emission. In particular, {\EP} events are detected in a broad cosmological range, extending to $z\approx5$, with a substantial fraction located at $z>2$ (see Figure~\ref{fig:zhist}).

With $z=5.47$, EP260119a extends the confirmed GQFXT population beyond $z\approx5$ and currently represents its furthest known member (Figure~\ref{fig:zhist}). Its redshift places it in the same regime as the most distant long GRBs detected by \textit{Swift}/BAT. Together with the similarity between the FXT and long-GRB redshift distributions, this suggests that GQFXTs trace massive-star formation over a comparable range of cosmic history.

Although EP260119a extends the GQFXT population to higher $z$, deriving a volumetric event rate is beyond the scope of this work. Such an estimate requires a well-characterised \textit{Einstein Probe} selection function and a substantially larger sample of securely classified GQFXTs. We therefore defer a quantitative population analysis to future studies.

Our prompt gamma-ray constraints, together with the inferred relativistic afterglow, indicate that the absence of detected gamma-ray emission does not imply an intrinsically weak explosion. Instead, it is consistent with scenarios in which the prompt gamma-ray emission is intrinsically soft, falls below the sensitivity of current gamma-ray instruments, or is reduced by viewing-angle effects \citep{Ricci2025,Li2026b}. As the \textit{Einstein Probe} sample continues to grow, it will become possible to determine whether GQFXTs simply represent the softest extension of the long-GRB population or constitute a distinct class of relativistic explosions, and to quantify their relative abundance with respect to classical gamma-ray-selected events.

\subsection{Transient sky at z > 5}

At redshifts $z > 5$, the observable transient sky is strongly shaped by instrumental selection effects \citep{Coward2013}. Classical high-$z$ GRBs have historically been identified through prompt gamma-ray triggers followed by rapid optical/near-infrared follow-up observations \citep{Salvaterra2009,Cucchiara2011}. Nevertheless, this strategy naturally favours events with bright hard X-ray and gamma-ray prompt emission, potentially biasing the observed population toward the most luminous relativistic explosions.

EP260119a can be compared with GRB~240315C/EP240315a ($z=4.859$) \citep{Liu2025,Gillanders2024,Levan2025}, considered the prototype of the emerging high-redshift FXT population discovered by \textit{Einstein Probe}. Both events were identified through bright and long-lasting soft X-ray emission detected by {\EP}/WXT, exhibited luminous optical counterparts, and showed unusually weak prompt gamma-ray emission compared with classical long GRBs. In the case of EP240315a, the gamma-ray signal was recovered only through offline analyses of \textit{Swift}/BAT and \textit{Konus}/Wind data, whereas EP260119a remained undetected by multiple gamma-ray instruments despite extensive searches (Section~\ref{sec:gammaray}). Together, these events suggest that \textit{Einstein Probe} is unveiling a population of distant relativistic transients whose prompt high-energy emission is dominated by soft X-rays and which are underrepresented in classical gamma-ray-selected samples.

\begin{figure}
    \includegraphics[width=\linewidth]{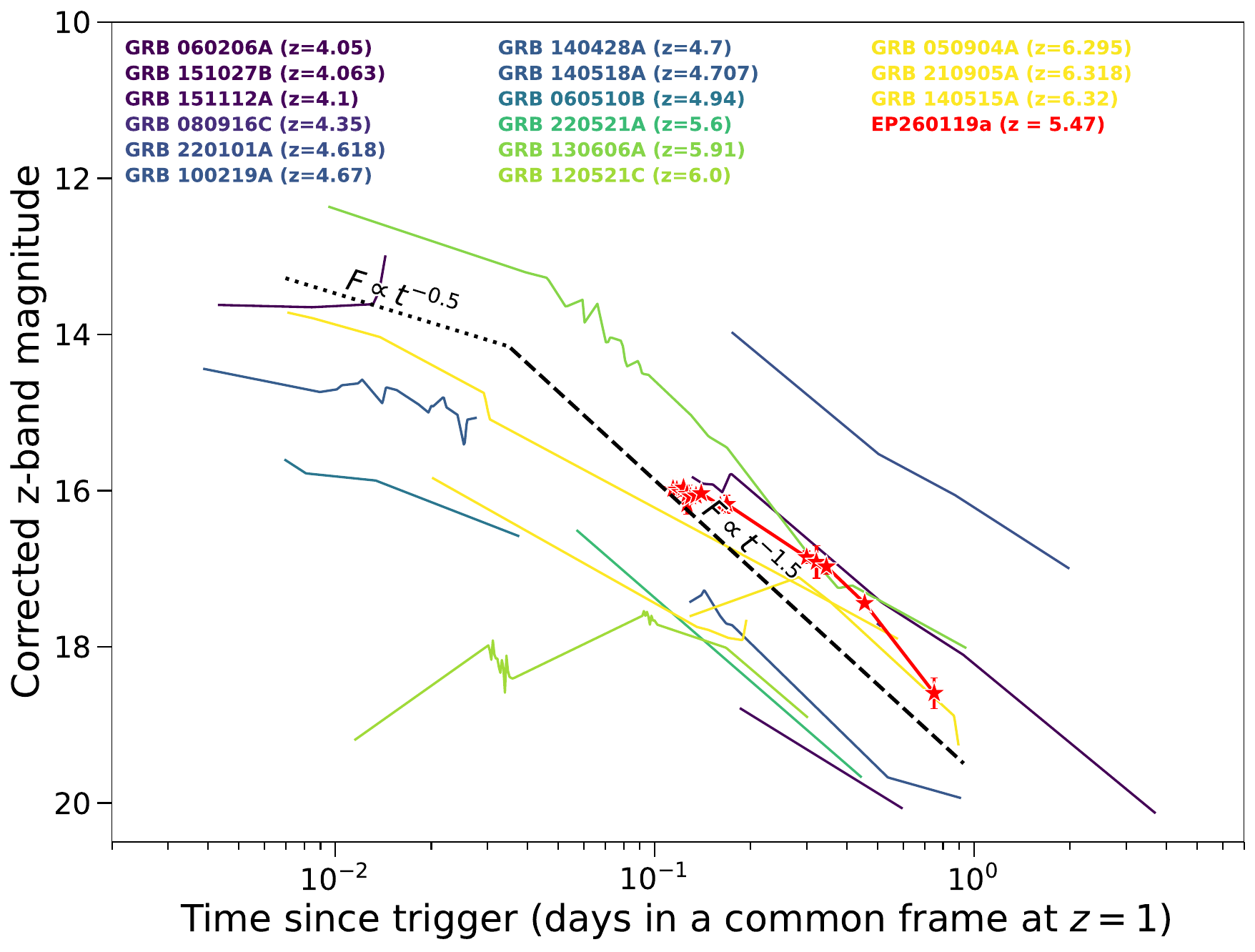}
    \caption{Light curves in the $z$ band of 15 GRBs with redshift $z>4$ taken from the Gamma-Ray Bursts Optical Afterglow Repository. We highlight EP260119a with star markers. We place the extinction corrected light curves in the $z=1$ system following the method described in \citet{Kann2006}, assuming a value of $\beta = -0.7$. We show two lines with temporal decaying index of $\alpha=-0.5$ and $\alpha=-1.5$ as reference. \label{fig:highz}}
\end{figure}

To place EP260119a in the context of the high-$z$ transient population, we compare its optical light curve with those of high-$z$ GRBs from the Gamma-Ray Bursts Optical Afterglow Repository \footnote{\url{https://grblc-catalog.streamlit.app}} \citep{Dainotti2024} in Figure~\ref{fig:highz}. Our subsample includes 15 events with $z > 4$ observed in the $z$ band (see Table~\ref{tab:highz-grbs}). 

After correcting the light curves to a common redshift following the methodology of \citet{Kann2006}, EP260119a falls well within the luminosity distribution of high-$z$ GRB afterglows. Its early optical evolution is comparable to that of several events in the sample, while the subsequent steep decline is also consistent with the diversity of temporal behaviours observed among high-$z$ GRBs. Therefore, although EP260119a is gamma-ray quiet, its optical afterglow does not appear exceptional when compared with the known population of distant GRBs, suggesting that the principal observational differences arise during the prompt high-energy phase rather than the afterglow evolution.

\begin{table}
\centering
\caption{Sample of GRBs with $z > 4$ and observations in the $z$ band obtained
from the database presented by \citet{Dainotti2024}.
References: (1) \citet{Fynbo2006}; (2) \citet{Selsing2019}; (3) \citet{18603};
(4) \citet{Greiner2009}; (5) \citet{Zhu2023}; (6) \citet{Thone2013};
(7) \citet{16181}; (8) \citet{16191}; (9) \citet{Dainotti2020};
(10) \citet{Dainotti2022}; (11) \citet{Price2007}; (12) \citet{32079};
(13) \citet{Hartoog2015}; (14) \citet{Laskar2014}; (15) \citet{Kawai2006};
(16) \citet{Saccardi2023}; (17) \citet{Fausey2025}; (18) \citet{Chornock2014};
(19) \citet{Melandri2015}; (20) \citet{Hjorth2012}.}
\label{tab:highz-grbs}
\begin{tabular}{llr}
\toprule
Event & Redshift & Reference \\
\midrule
GRB~060206A & 4.048 & 1,20 \\
GRB~151027B & 4.063 & 2 \\
GRB~151112A & 4.100 & 3 \\
GRB~080916C & 4.350 & 4 \\
GRB~220101A & 4.618 & 5 \\
GRB~100219A & 4.667 & 6, 20 \\
GRB~140428A & 4.700 & 7, 8 \\
GRB~140518A & 4.707 & 9, 10 \\
GRB~060510B & 4.940 & 11, 20 \\
GRB~220521A & 5.600 & 12 \\
GRB~130606A & 5.913 & 13 \\
GRB~120521C & 6.000 & 14 \\
GRB~050904A & 6.295 & 15 \\
GRB~210905A & 6.318 & 16, 17 \\
GRB~140515A & 6.320 & 18, 19 \\
\bottomrule
\end{tabular}
\end{table}

The extreme redshift of EP260119a also raises the possibility that some high-$z$ FXTs may originate from progenitor channels connected to the final stages of Population~III star formation \citep[see][for a recent review]{Klessen2023}, as previously proposed for high-$z$ GRBs \citep{Bromm2006}. At $z=5.47$, EP260119a lies within the epoch where such progenitors may still contribute to the transient population \citep{Morales-Rivera+2026}. However, there is currently no observational evidence linking EP260119a itself to a Population~III progenitor.

More broadly, EP260119a illustrates how wide-field soft X-ray surveys are beginning to reveal a population of distant relativistic explosions that is poorly represented in gamma-ray-selected samples. As the sample of FXTs discovered by {\EP} grows, the sample will provide a more complete census of relativistic transients during the first billion years of cosmic history.

\section{Conclusions}
\label{sec:summary}

We have presented the discovery and a comprehensive multiwavelength analysis of EP260119a, a GQFXT transient detected by \textit{Einstein Probe} whose bright optical counterpart was identified by COLIBRÍ. Subsequent spectroscopy established a redshift of $z=5.47$, making EP260119a the highest-$z$ FXT currently known.

The temporal and spectral evolution are consistent with synchrotron emission from a relativistic jet propagating into a stratified medium with a shallower density profile than a stellar wind. Assuming that the prompt emission can be described by a Band spectrum, we find that physically plausible spectral solutions consistent with the Einstein Probe flux and the available gamma-ray upper limits imply prompt energetics within the range commonly observed for long GRBs. Owing to the absence of a direct measurement of the prompt spectrum, these values should be regarded as illustrative rather than uniquely determined. This might suggest that EP260119a is not intrinsically under-energetic but instead represents a relativistic explosion whose prompt emission falls largely outside the sensitivity of current gamma-ray instruments.

Deep late-time GTC/HIPERCAM imaging revealed a faint extended source $1.11\pm0.42$\arcsec from the optical counterpart position. However, its photometric redshift is inconsistent with the spectroscopic redshift of EP260119a, and its relatively large chance-coincidence probability suggests that its association with EP260119a remains uncertain.

EP260119a demonstrates that gamma-ray quiet relativistic explosions can already be detected within the first billion years of cosmic history. Its broadband evolution is well described by synchrotron emission from a relativistic jet propagating into a nearly constant-density, shallowly stratified medium while the non-detections by \textit{SVOM}/ECLAIRs, \textit{SVOM}/GRM, \textit{Swift}/BAT, and \textit{Konus}-Wind suggest that the principal observational differences with classical GRBs arise during the prompt high-energy phase rather than during the afterglow.

These results highlight the complementary capabilities of current time-domain facilities. While \textit{Einstein Probe} has opened a new observational window onto the high-$z$ transient Universe through its sensitivity to soft X-ray emission, rapid-response ground-based facilities such as COLIBRÍ are essential for identifying optical counterparts, measuring their photometric evolution, and enabling the spectroscopic observations required to establish their distances and physical nature. Together, these capabilities suggest that relativistic explosions with prompt emission dominated by soft X-rays may be systematically under-represented in traditional gamma-ray-selected samples.

A more complete view of relativistic explosions in the early Universe will therefore require the synergy between sensitive wide-field soft X-ray surveys, rapid multiwavelength follow-up, and gamma-ray missions. As the \textit{Einstein Probe} sample continues to grow, it will become possible to quantify the extent of this selection bias and determine whether gamma-ray quiet FXTs represent the softest extension of the GRB population or a distinct class of relativistic explosions.

\section*{Acknowledgements}

COLIBRI received support from the French government under the France 2030 investment plan, as part of the Initiative d’Excellence d’Aix-Marseille Université-A*MIDEX through (ANR-11-LABX-0060 - OCEVU) and (AMX-19-IET-008 - IPhU), from LabEx FOCUS (ANR-11-LABX-0013), From Centre National d'Etudes Spatiale (CNES) and from CSAA-INSU-CNRS support program, and in Mexico from UNAM (Secretaria Administrativa, Coordinacion de la Investigacion Cientıfica, Instituto de Astronomıa and PAPIIT grant IN105921), and SECIHTI/CONACyT (277901, Ciencias de Frontera 1046632 and Laboratorios Nacionales).
The COLIBRÍ team thanks the staff of the Observatorio Astronómico Nacional at Sierra de San Pedro Mártir, as well as the technical and engineering teams at CEA, CPPM, IRAP, LAM, OHP, OSU Pytheas, and UNAM.

Data used in our work is taken from the catalogue \cite{Dainotti2024}, and the original data sources are cited within.

CAV acknowledges support from a SECIHTI fellowship. 

EAR acknowledges support from the UNAM/DGAPA Elisa Acuña and SECIHTI postdoctoral fellowships.

AK is supported by the UNAM/DGAPA Elisa Acuña postdoctoral fellowship.

MAA acknowledges support from MCIN/AEI/10.13039/501100011033 and the European Union grants PID2021-127495NB-I00 and PID2025-171322NB-C22 funded by, as well as from the Generalitat Valenciana through the Prometeo excellence programme grant CIPROM/2022/13. 

AMW is grateful for support from UNAM/DGAPA project IN109224.

BS acknowledges the support of the French Agence Nationale de la Recherche (ANR), under grant ANR-23-CE31-0011 (project PEGaSUS).

NG and LGG acknowledge the support of the Simons Foundation (MP-SCMPS-00001470, N. G., L. G. G.)

SR acknowledges the ACME project, which has received funding from the European Union’s Horizon Europe Research and Innovation program under Grant Agreement No. 101131928.

\section*{Data Availability}

The data underlying this article will be shared upon reasonable request to the corresponding author.



\bibliographystyle{mnras}
\bibliography{references} 




\appendix

\section{COLIBRÍ Pipeline}
\label{app:colibriasu}

COLIBRÍ ASU pipeline (Butler et al. in prep) is based on the ones developed for RATIR \footnote{\url{https://ratir.astroscu.unam.mx/}}, DDOTI \footnote{\url{https://ddoti.astroscu.unam.mx/}} and COATLI \footnote{\url{https://coatli.astroscu.unam.mx/}}. It automatically performs bias subtraction and flat-field correction, followed by astrometric calibration using the astrometry.net software \citep{astrometry}, iterative sky-subtraction, coaddition using SWARP \citep{swarp}, and source detection using SEXTRACTOR \citep{sextractor}.  

We calibrate our photometry against the PanSTARRS DR1 catalog \citep{Magnier2020}. Our systematic calibration error is about 1\%.

\section{Band function and Amati relation}
\label{appendix:band-amati}

To constrain the spectral properties and expected flux of the prompt emission consistent with the observation energy range of EP/WXT (i.e., $0.5\text{--}4.0\,\rm keV$), we model the observed photon spectrum using the empirical Band function \citep{Band+93}:
\begin{equation}\label{eq:Band_function}
    f_{\rm Band} (E_{\rm obs}) = A \,
    \begin{cases}
        \left( \frac{E_{\rm obs}}{100 \, \rm keV} \right)^\alpha \, \exp\left(-\frac{E_{\rm obs}}{E_0}\right), & E_{\rm obs} \leq (\alpha-\beta)E_0 \, , \\
        \left[ \frac{(\alpha-\beta)E_0}{100 \, \rm keV} \right]^{\alpha-\beta} \exp(\beta- \alpha) \left( \frac{E_{\rm obs}}{100 \, \rm keV} \right)^\beta, & E_{\rm obs} \geq (\alpha-\beta)E_0 \, ,
    \end{cases}
\end{equation}
where $A$ is the normalization factor in units of $\mathrm{ph \, cm^{-2} \, s^{-1} \, keV^{-1}}$, $\alpha$ and $\beta$ are the low- and high-energy photon spectral indices, respectively, and $E_0$ is the characteristic energy in the observer-frame. The characteristic energy is related to the intrinsic peak energy in the $\nu F_\nu \propto E_{\rm obs}^2 f_{\rm Band}(E_{\rm obs})$ spectrum as $E_{\rm peak, obs} =  (2 + \alpha) E_0 = E_{\rm peak}/(1+z)$.

We estimate $E_{\gamma, \rm iso}$ inverting the Amati relation \citep[$E_{\rm peak}/{\rm keV} = K ( E_{\gamma, \rm iso}/10^{52} {\rm erg})^m$][]{Amati+2002}, obtaining:
\begin{equation}\label{eq:inverse_amati}
    \log_{10} \left( \frac{E_{\gamma, \rm iso}}{10^{52} \, \rm erg} \right) = \frac{\log_{10} \left(E_{\rm peak}/\rm keV \right) - \log_{10} K}{m} \, ,
\end{equation}
with empirical parameters $K = 81.0 \pm 2.0$ and $m = 0.57 \pm 0.02$ \citep[see][]{Amati-06}. The isotropic equivalent energy $E_{\gamma, \rm iso}$ used in the relation of \cite{Amati+2002} is calculated integrating over the rest-frame $1\text{--}10\,000\,\rm keV$ energy band:
\begin{equation}\label{eq:Eiso_int}
    E_{\gamma, \rm iso} = \frac{4\pi d_L^2}{(1+z)} \int_{\frac{1 }{1+z}\, \rm keV}^{\frac{10\,000} {1+z} \, \rm keV} E_{\rm obs} \, N_{\rm Band} (E_{\rm obs}) \, dE_{\rm obs} \, ,
\end{equation}
where $d_L$ is the luminosity distance and $N_{\rm Band}$ is the time-integrated Band function over the burst duration. Here we assume that the photon spectral shape remains constant over the burst duration such that we can approximate it as $N_{\rm Band} \simeq T_{\rm 90, obs} f_{\rm Band} $, where $T_{90, \rm obs}$ is the observed duration of the burst containing 90\% of the detected fluence.

Therefore, Equations~\ref{eq:inverse_amati} and \ref{eq:Eiso_int} determine the normalization of the Band function as a function of $E_{\rm peak}$. And the observed EP/WXT flux can then be used to constrain $E_{\rm peak}$ through the relation
\begin{equation}\label{eq:EPWXT_flux}
F_{\rm EP/WXT} =
\int_{0.5\, \rm keV}^{4\, \rm keV}
E_{\rm obs} \, f_{\rm Band}(E_{\rm obs}) \, dE_{\rm obs} \, .
\end{equation}

Furthermore, the total uncertainty in $E_{\rm iso}$,  propagated in logarithmic space, is calculated assuming first-order Gaussian error propagation, which is given by:

\begin{equation}\label{eq:error_propagation}
    \sigma_{ \log_{10} E_{\gamma, \rm iso} } = \frac{1}{m }
    \sqrt{ \sigma_{ \ln_{10} E_{\rm peak}}^2 + \frac{\sigma_K^2}{ \left[ K \log(10) \right]^2 }   + \left[ \log_{10}\left( \frac{E_{\rm peak}/{\rm keV}}{K } \right) \right]^2 \sigma_m^2}  \, .
\end{equation}

Here, $\sigma_K=2.0$ and $\sigma_m=0.02$ are the uncertainties on the Amati-relation parameters. On the other hand, $\sigma_{\log_{10}E_{\rm peak}}=0.15$ represents the extra-Poissonian logarithmic dispersion intrinsic to the correlation  \citep{Amati-06}. Since $E_{\rm peak}$ is assumed rather than measured in our calculation, no measurement uncertainty is assigned to $E_{\rm peak}$.

\clearpage

\section{MCMC configuration for light curve fitting}
\label{app:mcmc-lc}

To ensure that the MCMC sampler fully explored the viable parameter space without being artificially constrained, we adopted broad uniform priors for all fitted parameters.

\begin{figure}
    \centering
    \includegraphics[width=\linewidth]{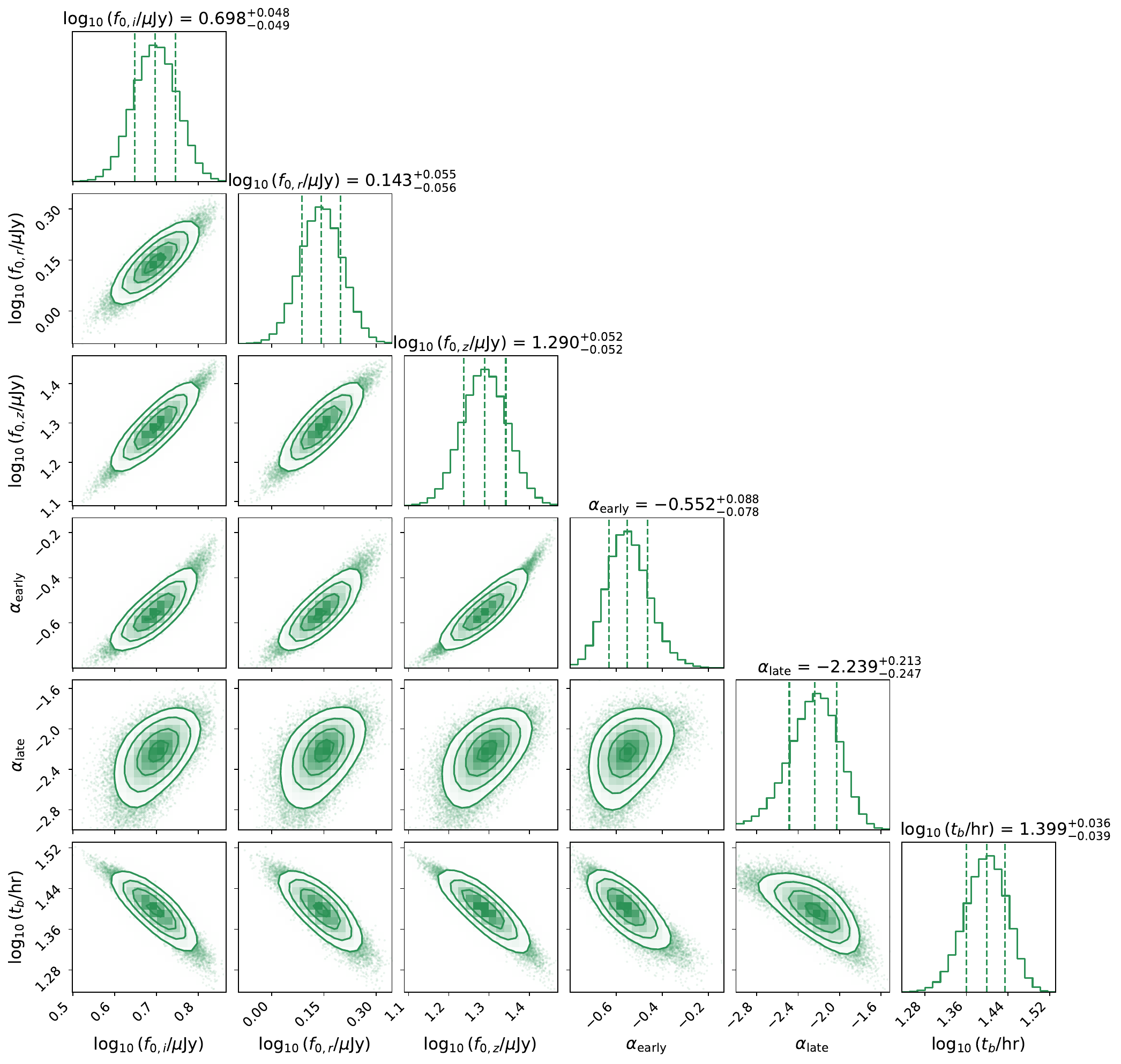}
    \caption{Corner plot showing the posterior parameter distributions for the smooth broken power-law fit to the multi-band afterglow light curve.}
    \label{fig:corner-plot-lc}
\end{figure}

\begin{figure*}
    \centering
    \includegraphics[width=1.0\linewidth]{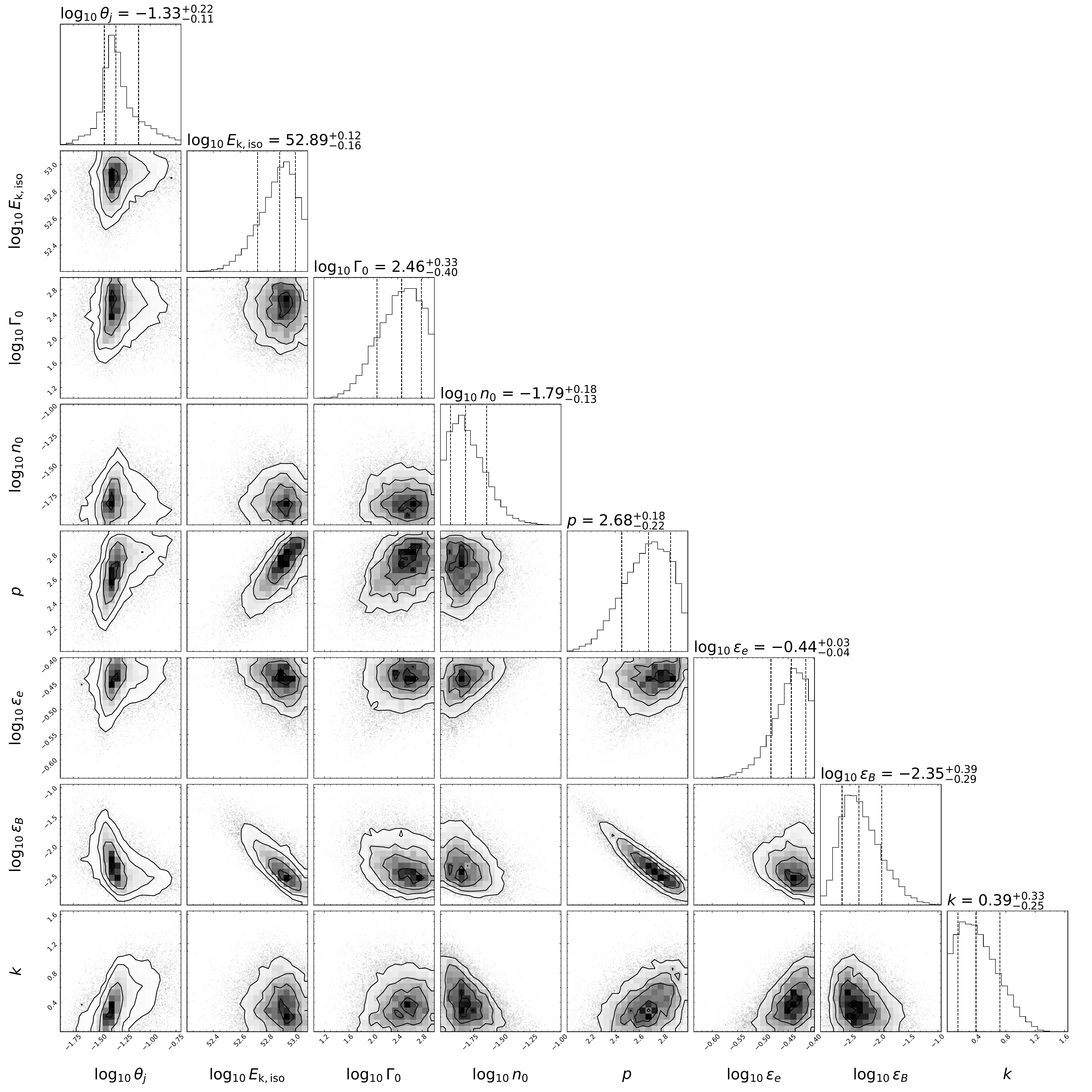}
    \caption{Model parameter posterior distributions obtained from an MCMC light curve fit.}
    \label{fig:corner_plots_top_hat_jet}
\end{figure*}

The allowed ranges for the smooth broken power-law parameters and band normalization ratios are summarized in Table~\ref{tab:mcmc_priors}. The results of the MCMC sampling are shown in Figure~\ref{fig:corner-plot-lc}. For an MCMC fit using a uniform jet model, we also use uniform priors. These values of the priors and posteriors are shown in Table~\ref{tab:mcmc_priors_uniform_jet} and Figure~\ref{fig:corner_plots_top_hat_jet}, respectively.

\begin{table}
    \centering
    \caption{Prior ranges for the joint smooth broken power law fit.}
    \label{tab:mcmc_priors}
    \begin{tabular}{llc}
        \hline\hline
        Parameter & Description & Range \\
        \hline
        $\log_{10}(f_{0,i}~/~\mathrm{\mu Jy})$           & $i$-band normalisation      & $(-10,\; 2)$    \\
        $\log_{10}(f_{0,r}~/~\mathrm{\mu Jy})$           & $r$-band normalisation      & $(-10,\; 2)$    \\
        $\log_{10}(f_{0,z}~/~\mathrm{\mu Jy})$           & $z$-band normalisation      & $(-10,\; 2)$    \\
        $s$                                              & Smoothing parameter (fixed) & 4.5              \\
        $\alpha_{\mathrm{early}}$                        & Pre-break decay index       & $(-0.8,\; 0.0)$  \\
        $\alpha_{\mathrm{late}}$                         & Post-break decay index      & $(-3,\; -0.8)$   \\
        $\log_{10}(t_{\mathrm{break}}~/~\mathrm{hr})$    & Break time                  & $(0.8,\; 2.0)$   \\
        \hline   
    \end{tabular}
\end{table}

\begin{table*}
    \centering
    \caption{Prior ranges for an MCMC fit using a uniform jet model. Uniform priors are adopted for all parameters.}
    \label{tab:mcmc_priors_uniform_jet}
    \begin{tabular}{llc}
        \hline\hline
        Parameter & Description & Range \\
        \hline
        $\log_{10}(\theta_{\rm jet}/{\rm rad})$ & Jet half opening angle & $(-2,\; -0.5)$    \\
        $\log_{10} (E_{k,\rm iso}/{\rm erg})$ & Isotropic-equivalent kinetic energy of the ejecta & $(51,\; 54)$ \\
        $\log_{10} \Gamma_{\rm 0}$ & Initial bulk Lorentz factor of the ejecta & $(1.5,\; 3)$ \\
        $\log_{10} (n_0/{\rm cm^{-3}})$ & Circumburst medium density at $R_0 = 10^{18}\, {\rm cm}$ & $(-2,\; 0.5)$ \\ 
        $p$ & Power-law index of accelerated electrons & $(2,\; 3)$ \\
        $\log_{10} \varepsilon_e$ & Energy fraction of accelerated electrons & $(-3,\; -0.5)$ \\ 
        $\log_{10} \varepsilon_B$ & Energy fraction of magnetic field & $(-6, -0.5)$ \\
        $k$ & Power-law index of an external medium density & $(0,\; 2)$ \\
        \hline   
    \end{tabular}
\end{table*}

\bsp	
\label{lastpage}
\end{document}